\documentclass[prb,aps,twocolumn,floats,nofootinbib,showpacs,superscriptaddress]{revtex4}

\usepackage{amsmath,amssymb,dsfont,graphicx,psfrag}
\usepackage{epsfig}
\usepackage{graphicx}

\def\ua{{\uparrow}}
\def\da{{\downarrow}}
\def\beq{\begin{equation}}
\def\eeq{\end{equation}}
\def\beqr{\begin{eqnarray}}
\def\eeqr{\end{eqnarray}}
\def\bq{{\vec q}}

\def\gperp{{g_{\perp}}}
\def\mO{{\mathbf O}}
\def\mQ{{\mathbf Q}}
\def\mS{{\mathbf S}}
\def\mH{{\mathbf H}}
\def\ma{{\mathbf a}}
\def\mb{{\mathbf b}}
\def\mc{{\mathbf c}}
\def\md{{\mathbf d}}
\def\mtau{{{\mathbf\tau}}}
\def\msigma{{{\mathbf\sigma}}}
\def\ttau{{\tilde{\mathbf\tau}}}
\def\tsigma{{\tilde{\mathbf\sigma}}}
\def\tV{{\tilde V}}
\def\ee{{\epsilon}}
\begin{document}
\title{Collective Bulk and Edge Modes through the Quantum Phase Transition in Graphene at $\nu=0$}

\author{Ganpathy Murthy}
\affiliation{Department of Physics and Astronomy, University of Kentucky, Lexington KY 40506-0055, USA}

\author{Efrat Shimshoni}
\affiliation{Department of Physics, Bar-Ilan University, Ramat-Gan 52900, Israel}

\author{H.~A.~Fertig}
\affiliation{Department of Physics, Indiana University, Bloomington, IN 47405, USA}

\date{\today}
\begin{abstract}
Undoped graphene in a strong, tilted magnetic field exhibits a radical
change in conduction upon changing the tilt-angle, which can be attributed to a
quantum phase transition from a canted antiferromagnetic (CAF) to a
ferromagnetic (FM) bulk state at filling factor $\nu=0$. This behavior
signifies a change in the nature of the collective ground state and
excitations across the transition.  Using the time-dependent
Hartree-Fock approximation, we study the collective neutral
(particle-hole) excitations in the two phases, both in the bulk and on
the edge of the system.  The CAF has gapless neutral modes in the
bulk, whereas the FM state supports only gapped modes in its bulk.  At
the edge, however, only the FM state supports gapless charge-carrying
states.  Linear response functions are computed to elucidate their
sensitivity to the various modes.  The response functions demonstrate
that the two phases can be distinguished by the evolution of a local
charge pulse at the edge.

\end{abstract}
\maketitle

\section{Introduction}\label{intro}
Graphene subject to a perpendicular magnetic field exhibits a quantum
Hall (QH) state at $\nu=0$. While such a state can exist in the
noninteracting model with a Zeeman coupling, the $\nu=0$ state in
experimental samples is believed to be driven by electron-electron
interactions
\cite{Zhang_Kim2006,Alicea2006,Goerbig2006,Gusynin2006,Nomura2006,Herbut2007,
 Jiang2007,Fuchs2007,Abanin2007,Checkelsky,Du2009,Goerbig2011,Dean2012,Yu2013}. In
such a hypothetical interacting state the bulk gap in the half-filled
zero Landau level would be associated with the formation of a
broken-symmetry many-body state. The variety of different ways to
spontaneously break the $SU(4)$ symmetry in spin and valley space
suggests an enormous number of potential ground states
\cite{Herbut2007AF,Jung2009,Nandkishore,Kharitonov_bulk,Roy2014,Lado2014,QHFMGexp}.

Recent experiments seem to see two such $\nu=0$
states\cite{Young2013,Maher2013}, with a phase transition between them tuned by
changing the Zeeman coupling strength. In these experiments, the
perpendicular field $B_{\perp}$ is kept fixed while the Zeeman
coupling is tuned by changing the parallel field
$B_{\parallel}$. Being at $\nu=0$, both states naturally show
$\sigma_{xy}=0$. At low Zeeman coupling, the sample has a vanishing
two-terminal conductance, indicating that its state is a ``vanilla''
insulator, whereas beyond a certain critical Zeeman coupling, the
sample has an almost perfect two-terminal conductance of $2e^2/h$,
suggesting that it is in a quantum spin  Hall state with protected edge states.

The most plausible interpretation of these experiments is in view of a
theoretical work of Kharitonov \cite{Kharitonov_bulk,bilayer_QHE_CAF},
which predicted
a $T=0$ quantum phase transition from a canted
antiferromagnetic (CAF) (the ``vanilla'' insulator) to a
spin-polarized ferromagnetic (FM) quantum Hall-like state tuned by
increasing the Zeeman energy $E_z$ to appreciable values. The behavior
of the two-terminal conductance is explained by the nature of the edge
modes\cite{Gusynin2008,Kharitonov_edge,Murthy2014,Pyatkovskiy2014,Knothe2015,Takei2015} of the two zero-temperature
phases. Previous investigations have shown that the FM state has a
fully gapped bulk, but supports gapless, helical, charged excitations
at its edge \cite{Abanin_2006,Fertig2006,SFP,Paramekanti}.
In analogy with the
quantum spin Hall (QSH) state in two-dimensional topological
insulators \cite{Kane-Mele,TIreview}, the gapless edge states of the
FM state are immune to backscattering by spin-conserving impurities
due to their helical nature: right and left movers have opposite spin
flavors.

While Kharitonov's proposal is consistent with the transport
experiments, there has been no direct experimental confirmation of the
nature of the two phases. In particlar, alternatives, such as a
Kekule-distorted phase\cite{Kekule}, are potential ground states for
the low-Zeeman ``vanilla'' insulator.

One of our motivations in this  study is to find physically measurable
quantities in both phases, both at the edge and in the bulk, that provide
characteristic signatures of each phase. In a previous
paper\cite{Murthy2014}, the present authors studied an extension of
Kharitonov's model (to include spin-stiffness) in the Hartree-Fock
(HF) approximation. We showed, using a simple model of the edge, that
a domain wall is formed near the edge.  This domain wall entangles the
spin and valley degrees of freedom, and leads to a single-particle
spectrum which is gapped everywhere.  In the bulk of the FM state, the
spins in both valleys are polarized along the total field, which we
will call $\uparrow$ for convenience. Deep into the edge, the state
must have vacuum quantum numbers, and so must be a singlet. The domain
wall is the region where the spin rotates continuously from being
fully polarized to being a singlet, thus acquiring an $XY$-component
in spin space. At the level of HF, this appears as a spontaneous
broken symmetry and an order parameter. Fluctuations about HF will
restore the symmetry in accordance with the Mermin-Wagner theorem.

In the FM phase, the low-energy charged excitations of the system
are gapless collective modes associated with a
$2\pi$ twist of the ground-state spin configuration in the
$XY$-plane\cite{Fertig2006,Murthy2014}. This spin twist is imposed upon the
position-dependent $S_z$ associated with the DW, thus creating a spin
{\it texture}, with an associated charge that is inherent to QH
ferromagnets \cite{QHFM,Fertig1994}.  Gapless 1D modes  associated with fluctuations of the DW
(which can be modeled as a helical Luttinger liquid \cite{SFP}) carry
charge and contribute to electric conduction.
In contrast, the CAF phase is characterized by a gap to charged
excitations on the edge \cite{Gusynin2008,Kharitonov_edge}, and a
broken $U(1)$ symmetry in the bulk (associated with $XY$-like order
parameter) implying a neutral, gapless bulk Goldstone mode.
As we have shown in our earlier work \cite{Murthy2014},  a proper description of the lowest energy charged excitations of
this state involves a coupling between  topological structures at
the edge and in the bulk, associated with the broken $U(1)$
symmetry.

In this paper we will carry forward our previous analysis, and focus
on the behavior of the collective particle-hole excitations in both
phases, which we compute in the time-dependent Hartree-Fock (TDHF)
approximation. Our goal is three-fold: Firstly, we want to verify that
the charged edge modes we proposed in previous work can be seen in
particle-hole excitations as well. Secondly, we will find
experimental signatures of the two different phases in the bulk as
well as at the edge. Thirdly, we want to compute a set of parameters
that we can use to build an effective theory of the edge.

The plan of this paper is as follows: In Section \ref{HFsec} we will
define our notational conventions and review the HF calculation of our
previous work. In Section \ref{TDHFsec} we will present the TDHF
formalism and general expressions for the spectral densities of
various correlation functions. In Section \ref{resultssec} we will present
our results, giving particular emphasis to the experimental signatures
of the bulk and edge collective modes. We end with conclusions and
discussion in Section \ref{conclusionsec}.

\section{Hamiltonian and Hartree-Fock Approximation}\label{HFsec}

We start with some notational conventions. In the $n=0$ Landau
level of graphene, there are two spin ($\ua$ and $\da$) and two valley
($K$ and $K'$) degrees of freedom. In Landau gauge, we can label the
orbital part of the single-particle states by $X=kl^2$, where $k$ is
the wavevector in the $y$-direction. We order our four-component
fermion destruction operators as
$\vec{\bf c}_k=({\bf c}_{K\ua,k},{\bf c}_{K\da,k},{\bf c}_{K'\ua,k},{\bf c}_{K'\da,k})^T$.
(Note that throughout this paper, operator quantities are indicated by boldface type.)
We define
Pauli matrices $\mathbf{\sigma}$ acting in the spin space and
$\mathbf{\tau}$ acting in the valley space
 with
$\sigma_0=\tau_0=I$ (the identity matrix), and
define $\ell=\sqrt{\frac{\hbar}{eB_{\perp}}}$ as the magnetic length. With these
notations  the Hamiltonian (first proposed by Kharitonov \cite{Kharitonov_bulk}) becomes
\begin{widetext}
\beq
\mH=\frac{\pi
  \ell^2}{L^2}\sum\limits_{k_1,k_2,\bq}\sum\limits_{a=0,x,y,z}e^{-\frac{(q\ell)^2}{2}}e^{i(\Phi(k_1,\bq)+\Phi(k_2,-\bq)}:g_a\vec{\mc}^{\,\dagger}_{k_1-q_y}\tau_a\vec{\mc}_{k_1}\vec{\mc}^{\,\dagger}_{k_2+q_y}\tau_a\vec{\mc}_{k_2}:-\sum_{k}
U_e(k)\vec{\mc}^{\dagger}_{k}\tau_x\vec{\mc}_{k} -E_Z\sum_{k}
\vec{\mc}^{\,\dagger}_{k}\sigma_z\vec{\mc}_{k},
\label{Hmicro}
\eeq
\end{widetext}
where $L$ is the linear size of the system and $\Phi(k,\bq)=\ell^2(-q_xk-\frac{1}{2}q_xq_y)$.

Note that the SU(4) symmetric $g_0$ term in the model does not affect
the groundstate phase of the system, and was not included in
Ref. \onlinecite{Kharitonov_bulk}.  It is added here to simulate the
spin/valley stiffness that we expect from the long-range Coulomb
interaction.  We have followed the common device of modelling the edge
as a smooth potential that couples to $\tau_x$, forcing the ground
state to be an eigenstate of $\tau_x$ deep inside the
edge. Furthermore, $g_x=g_y \equiv g_{xy}<0$, and $g_z>|g_{xy}|$ as
required for the system to be in the CAF or FM
groundstates. Throughout this paper we will present results for the
representative values $g_0=5$, $g_z=0.5$, $g_{xy}=-0.1$. We have
checked that other values do not qualitatively alter the results.

In previous work \cite{Murthy2014}, we carried out a numerical static HF study, allowing
all possible one-body expectation values \cite{Murthy2014}. The
results can be expressed as follows: The HF single-particle states in
the lowest Landau level (LLL) are entangled combinations of spin and
valley characterized by two angles we label $\psi_a$ and $\psi_b$. In
the bulk these angles are equal to each other and constant, but near
the edge they differ from each other and vary with $k$. The states may
be parameterized in the form
\beqr
|a\rangle=&\frac{1}{\sqrt{2}}\big(\cos{\frac{\psi_a}{2}},-\sin{\frac{\psi_a}{2}},\cos{\frac{\psi_a}{2}},\sin{\frac{\psi_a}{2}}\big)^T,\nonumber\\ |b\rangle=&\frac{1}{\sqrt{2}}\big(-\cos{\frac{\psi_b}{2}},\sin{\frac{\psi_b}{2}},\cos{\frac{\psi_b}{2}},\sin{\frac{\psi_b}{2}}\big)^T,\nonumber\\ |c\rangle=&\frac{1}{\sqrt{2}}\big(\sin{\frac{\psi_a}{2}},\cos{\frac{\psi_a}{2}},\sin{\frac{\psi_a}{2}},-\cos{\frac{\psi_a}{2}}\big)^T,\nonumber\\ |d\rangle=&\frac{1}{\sqrt{2}}\big(\sin{\frac{\psi_b}{2}},\cos{\frac{\psi_b}{2}},-\sin{\frac{\psi_b}{2}},\cos{\frac{\psi_b}{2}}\big)^T.
\eeqr

\begin{figure}[t]
\includegraphics[width=\linewidth]{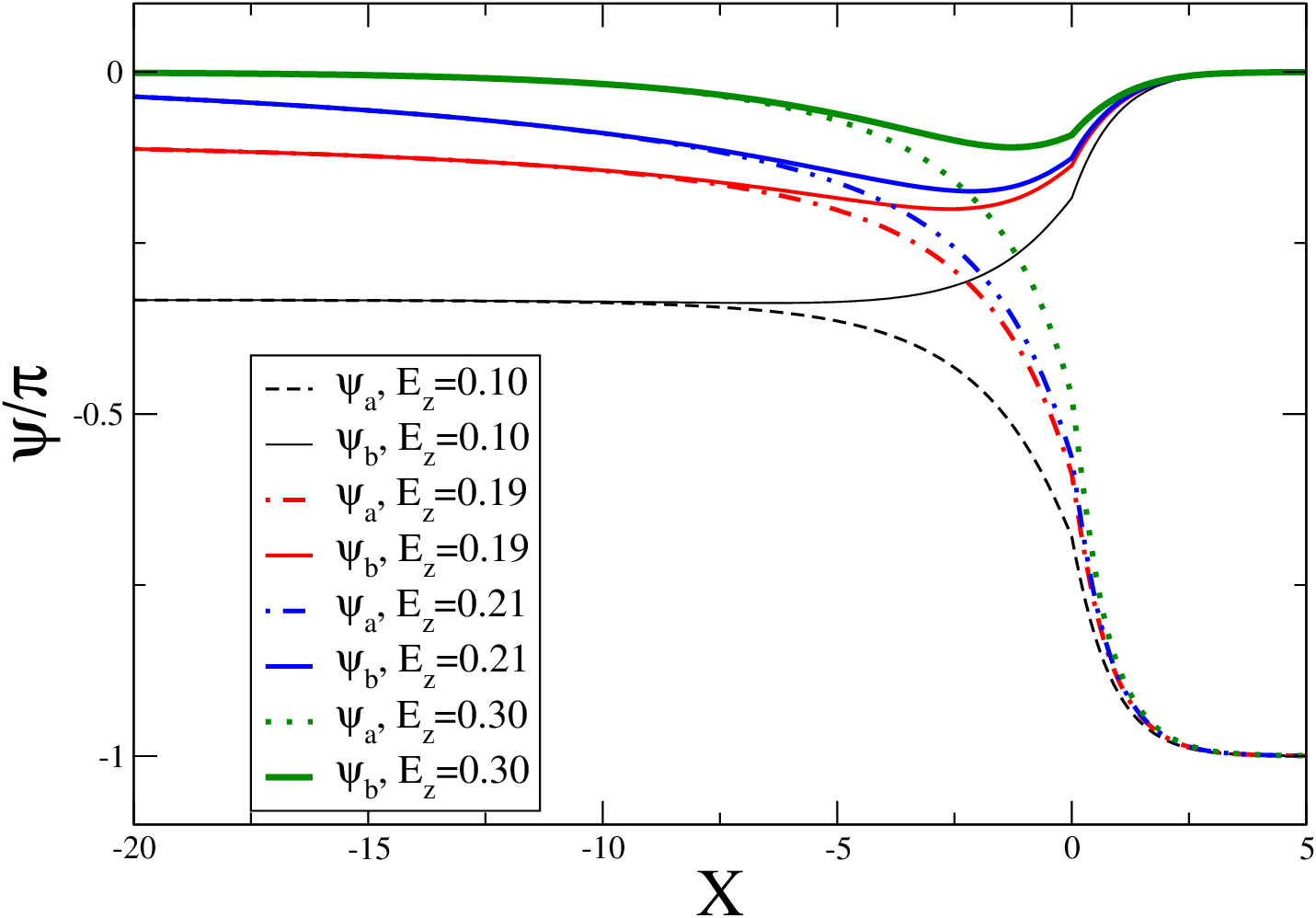}
\caption{(Color online.)  Variation of the canting angles $\psi_a$,
  $\psi_b$ with guiding center $X$ (in units of
  $\ell$), in the presence of an edge near $X=0$. The
  critical Zeeman energy is $E_z^c=0.2$.   }
\label{fig1}
\end{figure}

Defining $\gperp=|g_{xy}|$, in the bulk the values of
$\psi_a=\psi_b=\psi$ are given by $\cos{\psi}=\frac{E_Z}{2\gperp}$ for
$E_Z<E_{Zc}$, while $\psi=0$ for $E_Z>E_{Zc}$.
The quantum phase transition occurs at
$E_{Zc}=2\gperp=0.2$ in our units.
Fig. \ref{fig1} shows the variation of these angles as a function of
distance from the edge. The bulk is at negative values of $X=k\ell^2$, and the
edge potential linearly rises from $k=0$ to a maximum value of $U_e=5$
at $k\ell^2=3\ell$. In Fig. \ref{fig1} we have presented the angles
for four values of the Zeeman energy, two in the CAF phase and two in
the FM phase.

As the system approaches the transition from the FM side, there is a divergent length scale
\beq
\xi=\sqrt{\frac{g_0+g_z-3g_{xy}}{E_Z+2g_{xy}}},
\label{xibulk}\eeq
so that the edge effectively ``expands'' into the bulk. In the CAF phase, as
noted in the introduction, there is a spontaneously broken symmetry
(which can occur in two dimensional systems at $T=0$), which implies
the existence of a Goldstone mode \cite{Takei2015}. We will explicitly see this mode in
our TDHF calculations.

\begin{figure}[b]
\includegraphics[width=0.8\linewidth,trim={0 0 0 0.25cm},clip]{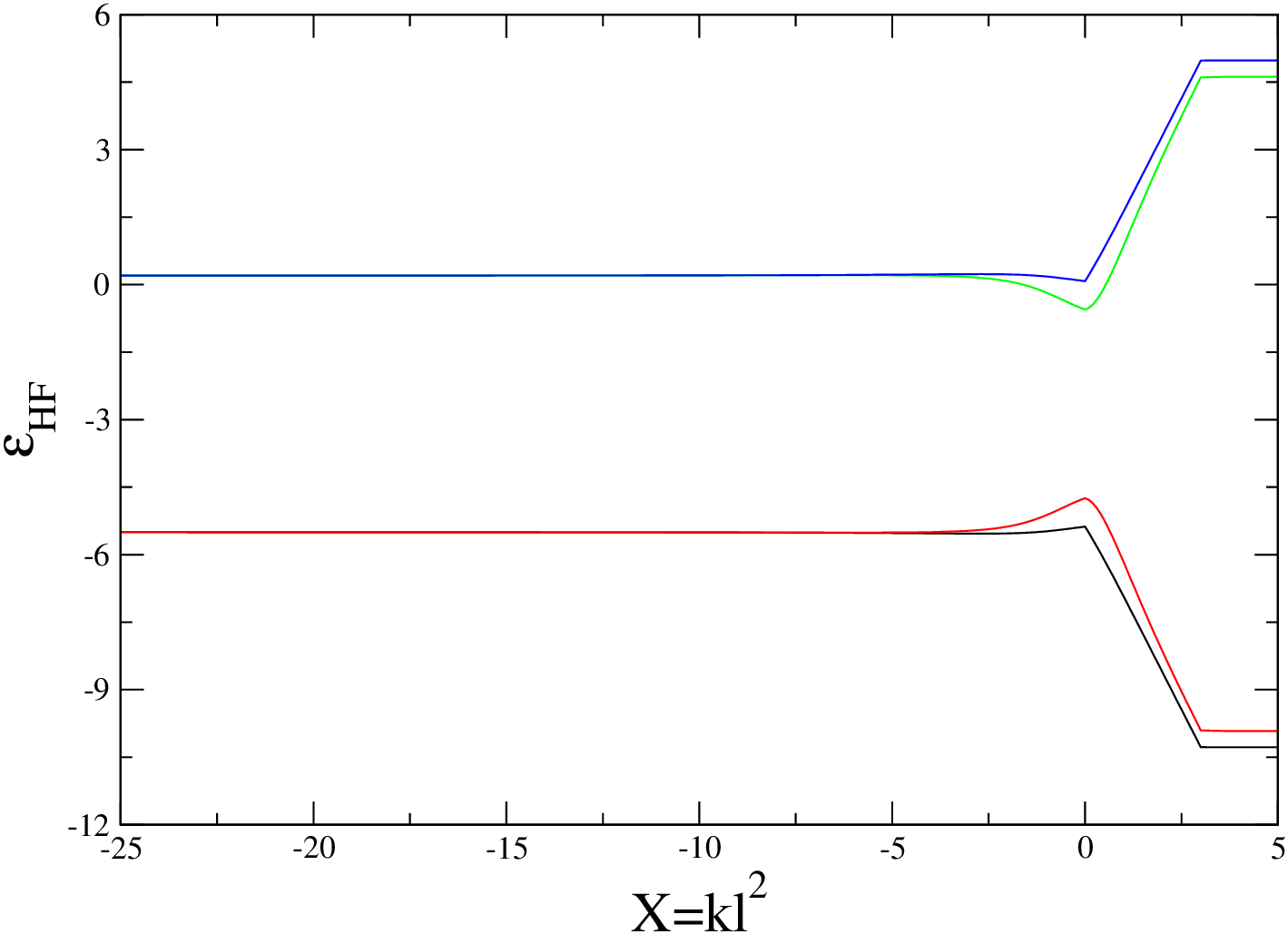}
\caption{(Color online.) Single-particle energies in the
  self-consistent HF state at $E_Z=0.22$ in the FM phase.}
\label{fig2}
\end{figure}

\begin{figure}[t]
\begin{center}
\includegraphics[width=0.8\linewidth,trim={0 0 0.15cm 0},clip]{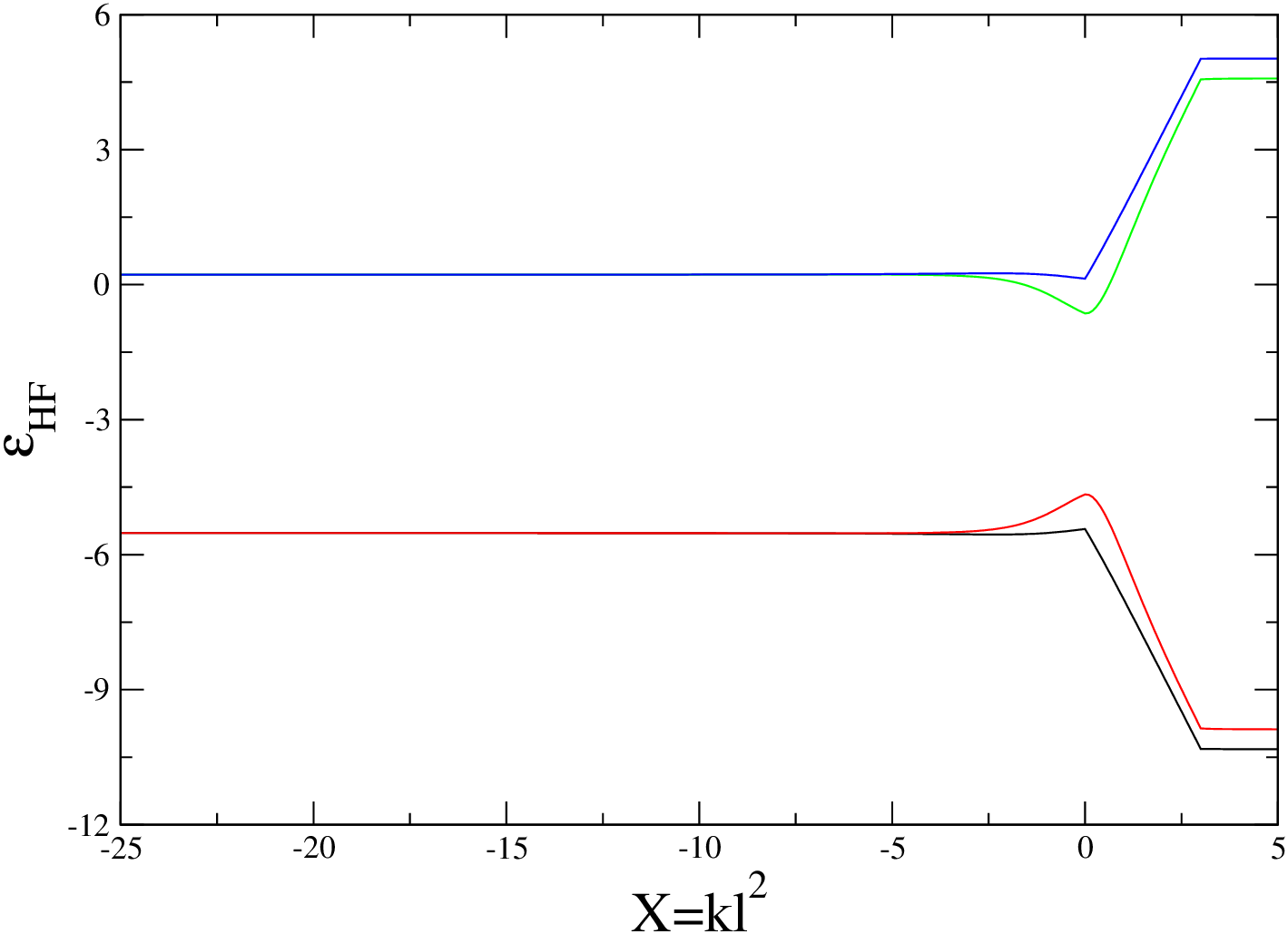}
\caption{(Color online.) Single-particle energies in the
  self-consistent HF state at $E_Z=0.18$, which is in the canted (CAF)
  phase. There is no closing of the single-particle gap.  }
\label{fig3}
\end{center}
\end{figure}

Fig. \ref{fig2} shows the single-particle
spectrum in the HF approximation for $E_Z=0.22$ on the FM side of the
transition, while Fig. \ref{fig3} shows it at $E_Z=0.18$, in the CAF
phase. There is no closing of the gap near the transition. This may
seem counterintuitive, especially on the FM side, where the
noninteracting model would predict a level crossing between states
carrying different spin quantum numbers. However, as noted before,
this is due to the spontaneous spin-mixing in HF. Naively, this would
indicate that the collective excitations will be gapped at the edge in
the FM phase. We will see how TDHF ``restores'' this symmetry and
predicts gapless edge excitations in the next section.

\vskip 0.25in
\vskip 0.25in

\section{Time-Dependent Hartree-Fock formalism}\label{TDHFsec}

The TDHF approximation consists of diagonalizing the Hamiltonian [Eq. (\ref{Hmicro})] in
the Hilbert space of particle-hole excitations. When used in
conjunction with the HF approximation, it is ``conserving''\cite{Kadanoff-Baym}, which
means that its results, though approximate, respect the symmetries of
the underlying Hamiltonian, {\it even if the HF solution breaks it}.

Let us briefly go through the TDHF for the bulk.

\subsection{Time-Dependent Hartree-Fock approximation in the bulk}
\label{TDHFbulksubsec}

The first step is to go to the basis in which the HF Hamiltonian is
diagonal. In the bulk this is independent of $k$. Let us call the
unitary matrix that carries out this basis change $U$. Explicitly, in
terms of $\psi_a,\ \psi_b$, we have

\beq
U=\left(\begin{array}{cccc}
                    \cos{\frac{\psi_a}{2}}&-\cos{\frac{\psi_b}{2}}&\sin{\frac{\psi_a}{2}}&\sin{\frac{\psi_b}{2}}\cr
-\sin{\frac{\psi_a}{2}}&\sin{\frac{\psi_b}{2}}&\cos{\frac{\psi_a}{2}}&\cos{\frac{\psi_b}{2}}\cr
\cos{\frac{\psi_a}{2}}&\cos{\frac{\psi_b}{2}}&\sin{\frac{\psi_a}{2}}&-\sin{\frac{\psi_b}{2}}\cr
\sin{\frac{\psi_a}{2}}&\sin{\frac{\psi_b}{2}}&-\cos{\frac{\psi_a}{2}}&\cos{\frac{\psi_b}{2}}\cr
\end{array}\right).
\label{Ugeneral}\eeq

We will refer to the four components of each operator $\vec{\mc}_k$ by
superscripts, as $\mc_k^i$. The subscript $k$ will be reserved for
labelling the Landau gauge wavefunctions. The new operators $\vec{\md}_{k}$
are related to the old ones $\vec{\mc}_{k}$ by

\beq
\mc_k^i=U_{ij}\md_k^j
\eeq

In order to re-express the Hamiltonian in terms of $\md^i_{k}$ it is
convenient to define matrices $\ttau_a$ and $\tsigma_a$ which are the matrices $\mtau$ and $\msigma$ unitarily transformed into
the basis of the $\vec{\md}_{k}$. Recalling that the angles $\psi_{a,b}=\psi$
are equal and constant we obtain
\beqr
\ttau_x=&U^{\dagger}\mtau_x U=\cos{\psi} \mtau_z\msigma_z+\sin{\psi} \mtau_x,\cr
\ttau_y=&\cos{\psi}\msigma_y-\sin{\psi} \mtau_y\msigma_x,\cr
\ttau_z=&-\mtau_z\msigma_x,\cr
\tsigma_z=&\cos{\psi} \mtau_z+\sin{\psi} \mtau_x\msigma_z.
\label{ttaubulk}\eeqr

Further defining
\beq
{\tilde V}_{ijlm}=\sum\limits_{a=0}^{3} g_a\big(\ttau_a\big)_{ij} \big(\ttau_a\big)_{lm},
\label{tVbulk}\eeq
we rewrite the Hamiltonian as
\begin{widetext}
\beq
\mH=\sum\limits_{k}-E_Z\md^{i\dagger}_{k}\md^j_{k}\big(\tsigma_z\big)_{ij} +\frac{\pi\ell^2}{L^2}\sum\limits_{kk'a\bq} e^{-iq_x(k-k')\ell^2-(q\ell)^2/2}e^{i[\Phi(k_1,\bq)+\Phi(k_2,-\bq)]} \md^{i\dagger}_{k-q_y/2}\md^{l\dagger}_{k'+q_y/2}\md^m_{k'-q_y/2}\md^j_{k+q_y/2} {\tilde V}_{ijlm}.
\eeq
\end{widetext}
The Hartree-Fock Hamiltonian  is obtained by reducing the two-body operators to one-body operators by using the expectation values
\beq
\langle \md^{i\,\dagger}_{k_1}\md^{j}_{k_2}\rangle=\delta_{k_1k_2}\delta_{ij} N_F(i),
\label{occbulk}\eeq
where $N_F(i)=0$ or $1$ is the occupation of the state $i$. We then have
\begin{widetext}
\beq
\mH_{HF}=\sum\limits_{k} \md^{m\,\dagger}_{k}\md^{n}_{k}\bigg(-E_Z\tsigma_{z,mn}+\sum\limits_{j}N_F(j)(\tV_{mnjj}-\tV_{mjjn})\bigg).
\eeq
\end{widetext}
In the self-consistent HF state, this one-body hamiltonian is diagonal in the $d$ basis with eigenvalues $\epsilon_m$, with $\epsilon_a=\epsilon_b$ and $\epsilon_c=\epsilon_d$.
Next, we define magnetoexciton operators \cite{Kallin1984} with  well-defined  momentum $\bq=(q_x,q_y)$ as
\beq
\mO_{mn}(\bq)=\frac{\sqrt{2\pi\ell^2}}{L}\sum\limits_{k} e^{-iq_xk\ell^2} \md^{m\dagger}_{k-q_y/2}\md^n_{k+q_y/2}.
\label{MEbulk}\eeq

One then takes the commutator $[\mH,\mO^{ij}_{k,q_y}]$ which will
contain both one-body and two-body terms; the latter are reduced
to one-body terms using HF expectation values.  After some algebra the final result is
\begin{widetext}
\beq
[\mH,\mO_{mn}]|_{HF}=(\epsilon_m-\epsilon_n)\mO_{mn}(\bq)+e^{-(q\ell)^2/2}(N_F(n)-N_F(m))\sum\limits_{ij}\big(\tV_{nmij}-\tV_{imnj}\big)\mO_{ij}(\bq).
\label{TDHFbulkeq}\eeq
\end{widetext}

It is clear that the magnetoexciton operators $\mO_{mn}$ for which
$N_F(m)=N_F(n)$ will propagate freely and will decouple from those
with $N_F(m) \ne N_F(n)$.
Thus, we can confine ourselves at each $\bq$ to a
set of 8 particle-hole operators, which we label by the following assignment to the
pair $(m,n)$: $(a,c)\to 1$, $(a,d)\to 2$, $(b,c)\to 3$, $(b,d)\to 4$,
$(d,b)\to 5$, $(c,b)\to 6$, $(d,a)\to 7$, and $(c,a)\to 8$. We will
identify the first four with bosonic destruction operators
$\ma_{\alpha}$ and the second four with creation operators
$\ma^{\dagger}(-\bq)$. It can easily be checked that, when HF averages are taken
on the right-hand side, the commutators satisfy bosonic relations

\beqr
[\ma_{\alpha}(\bq_1),\ma_{\beta}^{\dagger}(\bq_2)]|_{HF}=&\delta_{\alpha\beta}\delta_{\bq_1\bq_2},\cr
[\ma_{\alpha}(\bq_1),\ma_{\beta}(\bq_2)]|_{HF}=&0=[\ma^{\dagger}_{\alpha}(\bq_1),\ma^{\dagger}_{\beta}(\bq_2)]|_{HF}.
\eeqr

The FM phase is particularly simple since $\psi=0$. In this
state the creation and destruction operators do not mix. Defining the notations $\epsilon_0=\epsilon_c-\epsilon_a=2E_Z+g_0+g_z-2\gperp$ and $f(q)=e^{-(q\ell)^2/2}$ the
$4\times4$ matrix of the TDHF Hamiltonian in the subspace of destruction
operators is
\begin{widetext}
\beq
H^{(+)}_{TDHF}=\left(\begin{array}{cccc}
                   -\epsilon_0+f(q)(g_0+\gperp)&0&0&-f(q)(g_z+\gperp)\cr
                   0&-\epsilon_0+f(q)(g_0-\gperp)&-f(q)(g_z-\gperp)&0\cr
                   0&-f(q)(g_z-\gperp)&-\epsilon_0+f(q)(g_0-\gperp)&0\cr
                   -f(q)(g_z+\gperp)&0&0&-\epsilon_0+f(q)(g_0+\gperp)\cr
                   \end{array}\right).
\eeq
\end{widetext}
The TDHF Hamiltonian in the subspace of the creation operators is the
same as above, with an overall minus sign. Diagonalization is trivial,
leading to the (positive) eigenvalues

\beqr
\omega_1(q)=&2E_Z+(g_0+g_z)(1-f(q))-2\gperp(1+f(q)),\cr
\omega_2(q)=&2E_Z+(g_0+g_z-2\gperp)(1-f(q)),\cr
\omega_3(q)=& 2E_Z-2\gperp+g_0(1-f(q))+g_z(1+f(q)),
\eeqr
where the last mode is two-fold degenerate.  In the limit $q\to0$ we
see that the first mode has a gap of $\omega_1(0)=\Delta=2(E_Z-E_{Zc})$ where $E_{Zc}=2\gperp$.  This mode becomes
critical at the transition. The second mode has the limit
$\omega_2(0)=2E_Z$ and is the Larmor mode. Note that the Larmor
mode is unrenormalized by interactions, as expected from the translational
symmetry of the system.  This works out correctly even though the energy difference
between single-particle eigenstates of the static
HF Hamiltonian with opposite spin are interaction-dependent,
and is an example of how the TDHF approximation preserves symmetries
which may be broken in static HF\cite{Kadanoff-Baym}.

For the canted phase things are a bit more complicated. The
single-particle gap is
$\ee_0=\ee_c-\ee_a=2E_Z\cos{\psi}+g_0+g_z-2\gperp\cos{2\psi}$. The
creation and destruction subspaces do get mixed by the action of the
TDHF Hamiltonian. However, the matrix is block diagonal, with modes
$1,4,5,8$ mixing among themselves, while modes $2,3,6,7$ mix among
themselves separately. The $4\times4$ matrix in the $1,4,5,8$ subspace
is
\begin{widetext}
\beq
H_{TDHF}^{(1458)}=\left(\begin{array}{cccc}
                    -\ee_0+f(q)(g_0+\gperp)&-f(q)(g_z+\gperp \cos{2\psi})&2f(q)\gperp\sin^2{\psi}&0\cr
-f(q)(g_z+\gperp\cos{2\psi})&-\ee_0+f(q)(g_0+\gperp)&0&2f(q)\gperp\sin^2{\psi}\cr
-2f(q)\gperp\sin^2{\psi}&0&\ee_0-f(q)(g_0+\gperp)&f(q)(g_z+\gperp\cos{2\psi})\cr
0&-2f(q)\gperp\sin^2{\psi}&f(q)(g_z+\gperp\cos{2\psi})&\ee_0-f(q)(g_0+\gperp)\cr
\end{array}\right).
\eeq
This can also be easily diagonalized, with the (positive) eigenvalues being
\beqr
\omega_1(q)=&\sqrt{\big(2E_Z\cos{\psi}+(g_0+g_z)(1-f(q))-\gperp(f(q)+(2+f(q))\cos{2\psi})\big)^2-4\big(\gperp f(q)\sin^2{\psi}\big)^2},\cr
\omega_{2}(q)=&\sqrt{\big(2E_Z\cos{\psi}+g_0(1-f(q))+g_z(1+f(q))+\gperp(f(q)+(2-f(q))\cos{2\psi})\big)^2-4\big(\gperp f(q)\sin^2{\psi}\big)^2}.\cr
\eeqr
Similarly the TDHF matrix in the $2367$ block is
\beq
H_{TDHF}^{(2367)}=\left(\begin{array}{cccc}
                    -\ee_0+f(q)(g_0-\gperp\cos{2\psi})&-f(q)(g_z-\gperp)&-2f(q)\gperp\sin^2{\psi}&0\cr
-f(q)(g_z-\gperp)&-\ee_0+f(q)(g_0-\gperp\cos{2\psi})&0&-2f(q)\gperp\sin^2{\psi}\cr
2f(q)\gperp\sin^2{\psi}&0&\ee_0-f(q)(g_0-\gperp\cos{2\psi})&f(q)(g_z-\gperp)\cr
0&2f(q)\gperp\sin^2{\psi}&f(q)(g_z-\gperp)&\ee_0-f(q)(g_0-\gperp\cos{2\psi})\cr
\end{array}\right).
\eeq
One of the (positive) eigenvalues of this matrix is the same as $\omega_2(q)$ above, while the other is
\beq
\omega_3(q)=\sqrt{\big(2E_Z\cos{\psi}+(g_0+g_z)(1-f(q))-\gperp((2-f(q))\cos{2\psi}-f(q))\big)^2-4\big(f(q)\gperp\sin^2{\psi}\big)^2}.
\eeq
\end{widetext}

\begin{figure}[t]
\includegraphics[width=0.8\linewidth]{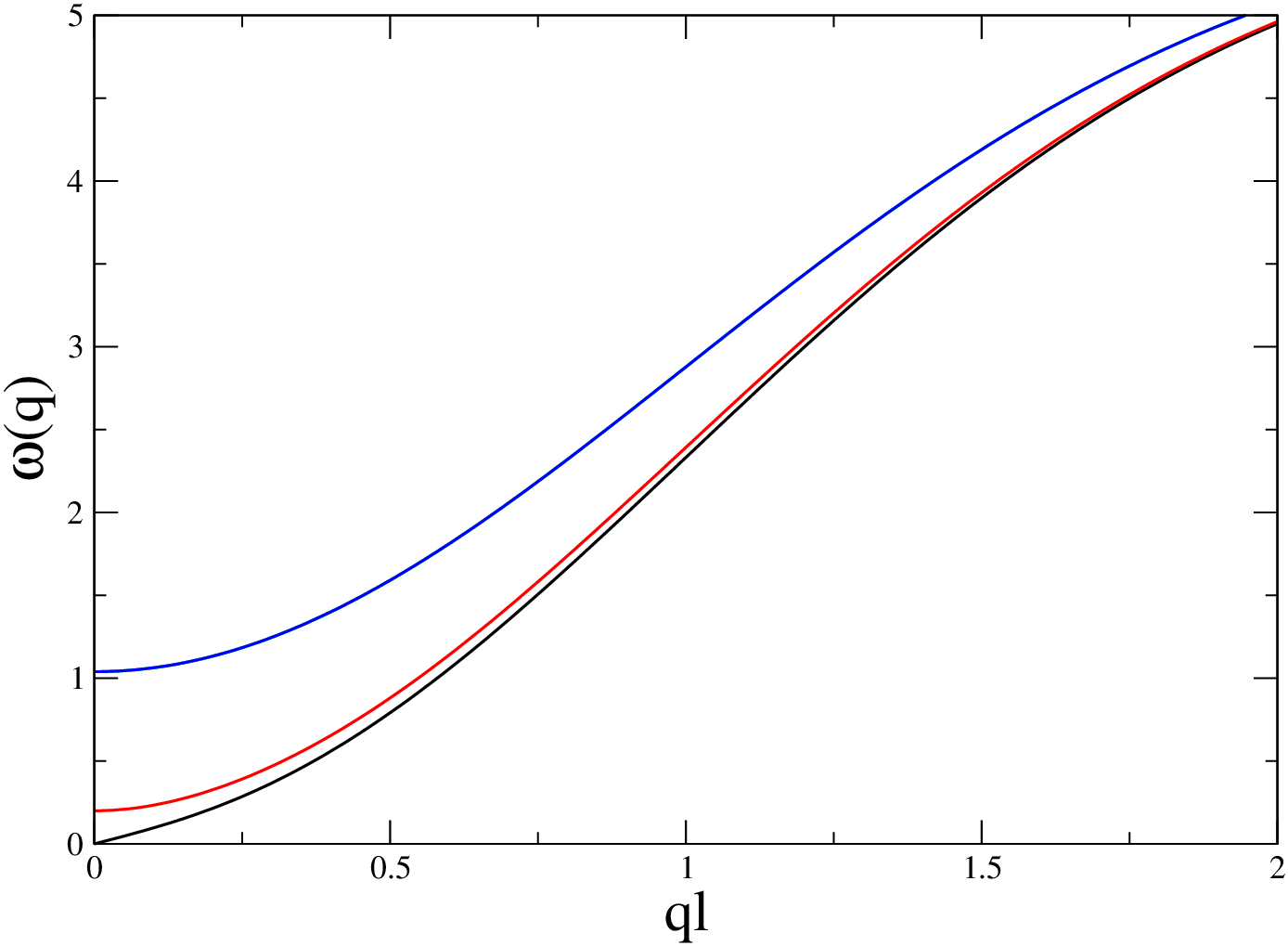}
\caption{(Color online.)  Bulk collective modes at $E_Z=0.1$, deep in
  the CAF phase. Note the linearly dispersing gapless mode which is
  the Goldstone mode of the broken $U(1)$ symmetry. }
\label{bulk-NM-ez0p10fig}
\end{figure}

\begin{figure}[b]
\includegraphics[width=0.8\linewidth]{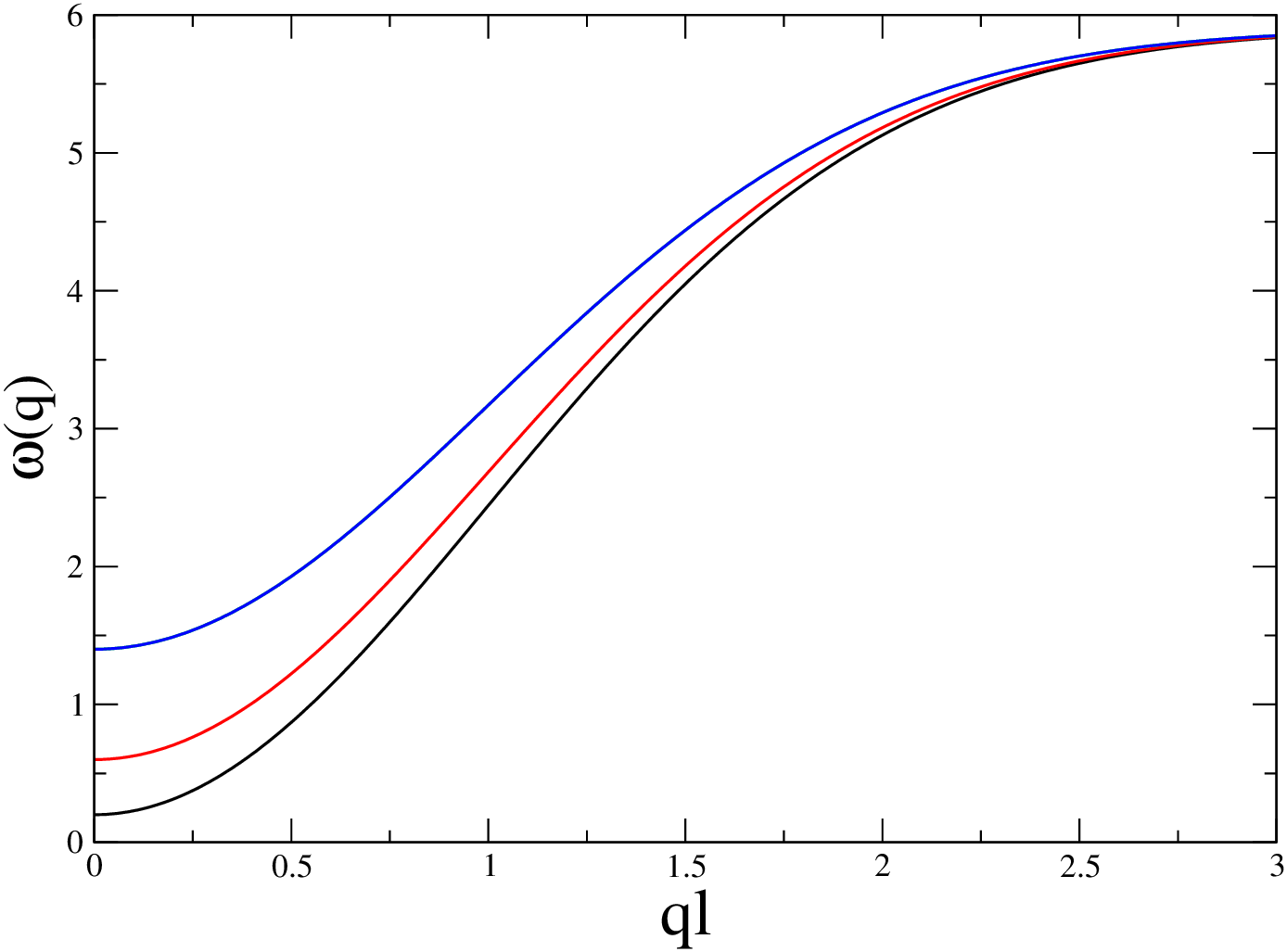}
\caption{(Color online.)  Bulk collective modes at $E_Z=0.3$, deep in
  the FM phase. All modes are robustly gapped. }
\label{bulk-NM-ez0p30fig}
\end{figure}

In the $q\to0$ limit $\omega_1(q)$ is the gapless Goldstone mode,
while $\omega_3(q)$ is the Larmor mode. The spin-wave velocity of
$\omega_1(q)$ can be extracted as
\beq
v_s=\ell\sqrt{2\gperp\sin^2{\psi}\big(g_0+g_z+\frac{E_Z^2}{2\gperp}\big)}.
\eeq
As the system approaches criticality from below, defining $\Delta=E_{Zc}-E_Z=2\gperp-E_Z$ we see that $v_s\simeq\sqrt{\Delta}$.
 Examples of the collective bulk modes for the CAF and FM phases are presented
in Figs. \ref{bulk-NM-ez0p10fig} and \ref{bulk-NM-ez0p30fig}.
\vskip 0.25in

\vskip 0.25in
We next turn to TDHF in the system with an edge, which is considerably more involved.

\subsection{Time-Dependent Hartree-Fock approximation with an edge}
\label{TDHFedgesubsec}

There are several complications in the system with an edge. Firstly,
there is translation invariance only in the $y$ direction, so only
$q_y$ is a good quantum number for excitations. Secondly, the unitary
transformation defined in Eq. (\ref{Ugeneral}) will be $k$-dependent.
Consequently,  the interaction matrix elements in the HF basis will also depend explicitly on $k$,
\begin{widetext}
\beq
\tV_{ijlm}(k_1,k_2,q_y)=\sum\limits_{a=0,x,y,z} g_a \big(U^{\dagger}(k_1-q_y)\tau_aU(k_1)\big)_{ij}\big(U^{\dagger}(k_2+q_y)\tau_aU(k_2)\big)_{lm}.
\eeq
The Hamiltonian in this basis is

\beq
\mH=-\sum\limits_{k}\vec{\md}^{\,\dagger}_kU^{\dagger}(k)\big(E_Z\sigma_z+U_e(k)\tau_x\big)U(k)\vec{\md}_k+\frac{\pi\ell^2}{L^2}\sum\limits_{a,k_1,k_2,\bq}
e^{-\frac{(q\ell)^2}{2}}e^{i(\Phi(k_1,\bq)+\Phi(k_2,-\bq))}\tV_{ijlm}(k_1,k_2,q_y)\md^{i\dagger}_{k_1-q_y}\md^{l\dagger}_{k_2+q_y}\md^m_{k_2}\md^j_{k_1}.
\eeq
\end{widetext}

As in the bulk, the next step is to define the magnetoexciton
operators in the $d$ basis. We will keep $L_y$ finite, so
that the quantum number $q_y=\frac{2\pi j_y}{L_y}$ is discrete, and write
 \beq
\mO^{ij}_{k}(q_y)=\md^{i\dagger}_{k-q_y}\md^j_{k}. \eeq
One then takes the commutator $[\mH,\mO^{ij}_{k}(q_y)]$ which will contain
both one-body and two-body terms. We again reduce the two-body terms to
one-body terms by using the HF expectation values. In the $d$-basis
Eq. (\ref{occbulk}) remains true independent of $k$; {\it i.e.},
\begin{figure}[t]
\includegraphics[width=0.8\linewidth]{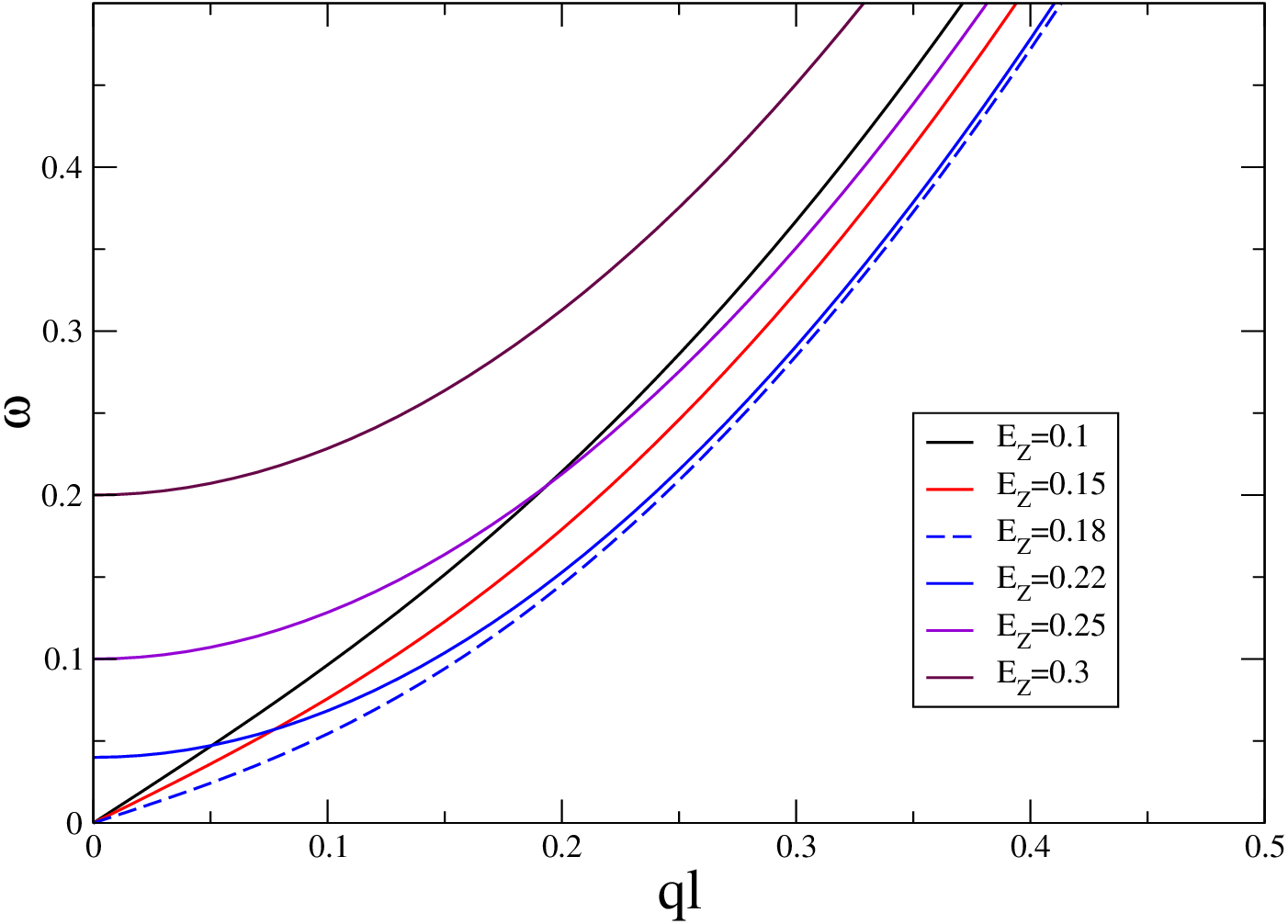}
\caption{(Color online.)  The lowest-lying bulk collective mode as it evolves with  $E_Z$. It can be seen that the mode becomes gapped at the transition, and the spin-wave velocity vanishes continuously as $E_Z\to E_{zc}^{-}$.   }
\label{comparisonfig}
\end{figure}
\vskip 0.25in
 \beq \langle
\md^{i\dagger}_{k_1}\md^j_{k_2}\rangle=\delta_{k_1k_2}\delta_{ij} N_F(i),
\eeq
where $N_F(i)=0$ or $1$ is the occupation of the state $i$. The
analog of Eq. (\ref{TDHFbulkeq}) in the system with an edge is
\begin{widetext}
\beqr
[\mH,\mO^{ij}_{k}(q_y)]_{HF}=&(\epsilon_{i}(k-q_y)-\epsilon_{j}(k))\mO^{ij}_{k}(q_y)+ \nonumber\cr \frac{N_F(j)-N_F(i)}{L_y\sqrt{2\pi\ell^2}}e^{-\frac{(q_y\ell)^2}{2}}
&\sum\limits_{k'}e^{-\frac{((k-k')\ell)^2}{2}} \mO^{lm}_{k'}(q_y)\big(\tV_{jilm}(k-q_y,k',-q_y)-\tV_{lijm}(k-q_y,k',k-k')\big).
\label{TDHFedgeeq}\eeqr
\end{widetext}
Thus, for every value of $q_y$ all the $\mO^{ij}_{k}(q_y)$ get
coupled to each other. This is a matrix diagonalization problem with
the dimension of the matrix being proportional to the number of $k$
values kept. As one approaches the transition, due to the diverging
length scale more $k$ values have to be retained.

As in the bulk,  one needs to consider only the operators which connect
filled with empty HF levels. Let us order the index $i=1,2,3,4$ in
order of increasing HF energy. Then $N_F(1)=N_F(2)=1$ and
$N_F(3)=N_F(4)=0$. It is convenient to divide the operators into two
groups: positive energy operators $\mO^{ij,(+)}_k(q_y)$ have $i<j$ while
negative energy operators $\mO^{ij,(-)}_k(q_y)$ have $i>j$. They are related by
\beq
\mO^{ij,(-)}_k(q_y)=\big(\mO^{ji,(+)}_{k+q_y}(-q_y)\big)^{\dagger}.
\eeq
To simplify the notation let us introduce a composite label for the
positive energy operators $\alpha={i,j,+,k}$ and the notation
\beq
\ma_{\alpha}(q_y)=\mO^{ij,(+)}_k(q_y).
\eeq
These operators share many features of canonical boson operators. In
particular, they satisfy canonical commutation relations upon taking a
HF average,
\beq
[\ma_{\alpha}(q_y),\ma^{\dagger}_{\beta}(q_y')]|_{HF}=\delta_{\alpha\beta}\delta_{q_yq_y'}.
\eeq
The TDHF equations can then be written as \beq
[\mH,\ma_{\alpha}(q_y)]=\sum\limits_{\beta}\big(A_{\alpha\beta}(q_y)\ma_{\beta}(q_y)+B_{\alpha\beta}(q_y)\ma^{\dagger}_{\beta}(-q_y)\big)
\eeq
and its adjoint. Note that these equations can be thought of as arising from
the Bogoliubov Hamiltonian
\beqr
\mH_B&=-\sum\limits_{\alpha\beta}\big(A_{\alpha\beta}(q_y)\ma^{\dagger}_{\alpha}(q_y)\ma_{\beta}(q_y)+h.c.\nonumber\cr
&+B_{\alpha\beta}(q_y)\ma^{\dagger}_{\alpha}(q_y)\ma^{\dagger}_{\beta}(-q_y)+h.c.\big).
\eeqr
Diagonalizing this Hamiltonian corresponds to finding eigenoperators
$\mb_{\mu}(q_y)$, $\mb^{\dagger}_{\mu}(-q_y)$ such that
\beqr
[\mH_B,\mb_{\mu}(q_y)]&= -E_{\mu}(q_y)\mb_{\mu}(q_y),\cr
 [\mH_B,\mb^{\dagger}_{\mu}(q_y)]&= E_{\mu}(q_y)\mb^{\dagger}_{\mu}(q_y).\cr
\eeqr
This makes it evident that the eigenvalues of $H_B$
come in $\pm$ pairs. The eigenoperators can be expressed in the original
$\alpha$ basis as
\beqr
\mb_{\mu}(q_y)=&\sum\limits_{\alpha}\big(\psi^{(+)}_{<,\mu\alpha}(q_y)\ma_{\alpha}(q_y)+\psi^{(-)}_{<,\mu\alpha}(q_y)\ma^{\dagger}_{\alpha}(-q_y)\big),\cr
\mb^{\dagger}_{\mu}(-q_y)=&\sum\limits_{\alpha}\big(\psi^{(+)}_{>,\mu\alpha}(q_y)\ma_{\alpha}(q_y)+\psi^{(-)}_{>,\mu\alpha}(q_y)\ma^{\dagger}_{\alpha}(-q_y)\big).
\label{b2a}
\eeqr
It is important to note that the orthonormalization of the
``wavefunctions'' $\psi^{(\pm)}_{\mu\alpha}$ is determined by the
commutation relation of the operators $b_{\mu},\ b^{\dagger}_{\mu}$
\beqr
[\mb_{\mu}(q_y),\mb^{\dagger}_{\nu}(q_y')]&=\delta_{\mu\nu}\delta_{q_yq_y'},\nonumber\cr
[\mb_{\mu}(q_y),\mb_{\nu}(q_y')]&=[\mb^{\dagger}_{\mu}(q_y),\mb^{\dagger}_{\nu}(q_y')]=0,
\eeqr
which imply

\beqr
&\sum\limits_{\alpha}\psi^{(+)}_{<,\mu\alpha}(q_y)\psi^{(+)*}_{<,\nu\alpha}(q_y')-\psi^{(-)}_{<,\mu\alpha}(q_y)\psi^{(-)*}_{<,\nu\alpha}(q_y')=\delta_{\mu\nu}\delta_{q_yq_y'},\cr
&\sum\limits_{\alpha}\psi^{(-)*}_{>,\mu\alpha}(q_y)\psi^{(-)}_{>,\nu\alpha}(q_y')-\psi^{(+)*}_{>,\mu\alpha}(q_y)\psi^{(+)}_{>,\nu\alpha}(q_y')=\delta_{\mu\nu}\delta_{q_yq_y'},\cr
&\sum\limits_{\alpha}\psi^{(+)}_{<,\mu\alpha}(q_y)\psi^{(+)*}_{>,\nu\alpha}(q_y')-\psi^{(-)}_{<,\mu\alpha}(q_y)\psi^{(-)*}_{>,\nu\alpha}(q_y')=0.
\eeqr
This provides us with a complete set of one-body operators in terms
of which any operator can be expanded, and can be exploited to find
linear response functions.

Consider a one-body operator
$\mQ(q_y)$. In the original basis we can expand it as
\beq
\mQ(q_y)=\sum\limits_{\alpha} Q^{(+)}_{\alpha}(q_y)\ma_{\alpha}(q_y)+Q^{(-)}_{\alpha}(q_y)\ma^{\dagger}_{\alpha}(q_y).
\label{expandQina}\eeq
 Employing Eq. (\ref{b2a}), we can also expand $Q$ in the eigenbasis of the TDHF Hamiltonian as
\beq
\mQ(q_y)=\sum\limits_{\mu} R^{(+)}_{\mu}(q_y)\mb_{\mu}(q_y)+R^{(-)}_{\mu}(q_y)\mb^{\dagger}_{\mu}(q_y).
\label{expandQinb}\eeq
To find the coefficients $R$ we simply take the commutator of $Q$ with
$\mb_{\mu},\ \mb^{\dagger}_{\mu}$, or alternatively use the
orthonormalization conditions, to obtain
\beqr
R^{(+)}_{\mu}=&\sum\limits_{\alpha}Q^{(+)}_{\alpha}\psi^{(+)*}_{<\mu\alpha}-Q^{(-)}_{\alpha}\psi^{(-)*}_{<\mu\alpha},\cr
R^{(-)}_{\mu}=-&\sum\limits_{\alpha}Q^{(+)}_{\alpha}\psi^{(+)*}_{>\mu\alpha}-Q^{(-)}_{\alpha}\psi^{(-)*}_{>\mu\alpha},
\eeqr
where we have suppressed the argument $q_y$ for compactness. Now the retarded $QQ$ response function can be written in the frequency domain as
\beq
\chi_{QQ}(q_y,\omega)=\sum\limits_{\mu}\bigg(\frac{|R^{(+)}_{\mu}(q_y)|^2}{\omega+i\eta+E_{\mu}(q_y)}-\frac{|R^{(-)}_{\mu}(q_y)|^2}{\omega+i\eta-E_{\mu}(q_y)}\bigg).
\eeq
Finally, the spectral density is defined by
\beq
S_{QQ}(q_y,\omega)=-\pi Im\big(\chi_{QQ}(q_y,\omega)\big).
\label{specden}\eeq

Since we are trying to find experimental signatures of the two phases,
we will focus on one-body operators that naturally couple to external
probes, which include the charge density and spin densities. When
computing response functions, we will assume that we are coupling the
relevant operator in a strip of width $\ell$. The perturbation
coupling to the density operator with $y$-momentum $q_y$, for example, will have the form
\begin{eqnarray}
\mQ_{\rho}(q_y)&=& \sum\limits_{k} e^{-(k-k_0)^2\ell^2/2}\vec{\mc}^{\dagger}_{k-q_y}\vec{\mc}_k  \\
&=&\sum\limits_{k}e^{-(k-k_0)^2\ell^2/2} \vec{\md}^{\,\dagger}_{k-q_y}U^{\dagger}(k-q_y)U(k)\vec{\md}_k \nonumber
\end{eqnarray}
When $k_0$ is near the edge, this will couple primarily to edge modes,
whereas if $k_0$ is deep in the bulk, it couples solely to bulk
modes. Deep in the bulk, since the angles $\psi_{a,b}$ are constant,
$U^{\dagger}(k-q_y)U(k)=1$, so $\mQ_{\rho}$ is diagonal in the
$\vec{\md}$ basis. Thus, there is no response to a density perturbation in the bulk.

Similar expressions for the spin-density operators are
\beq
\mS^a=\sum\limits_{k}e^{-(k-k_0)^2\ell^2/2} \vec{\md}^{\,\dagger}_{k-q_y}U^{\dagger}(k-q_y)\sigma_aU(k)\vec{\md}_k.
\eeq
Exactly as above, in the FM phase, there is no response to $\mS^z$.

We now proceed to the results.

\section{Results of the TDHF approximation}
\label{resultssec}

We will focus on correlators of interest, specifically the density-density,
$S_zS_z$, $S_{x}S_{x}$, and  $S_{y}S_{y}$  correlators. In each case we will plot the
spectral density of the correlator, the peaks of which will give us an
indication of the excitations that this correlator couples to, both for
the bulk and the edge.  The latter
will reveal the distinct character of the edge excitations.

In carrying
out TDHF for the bulk, one can use translation invariance to assume
that both $q_x$ and $q_y$ are good quantum numbers. This reduces the
problem to the diagonalization of an $8\times8$ matrix.
For the edge, we use a ``bulk'' of length $80\ell$ and an edge of
length $4\ell$. We choose $L_y=20\pi\ell$ so that the separation
between successive values of $k\ell^2$ is $0.1\ell$. This set of
parameters leads to the a TDHF matrix of dimension roughly $7000\times
7000$.  The finite size of the bulk means that we cannot approach the
phase transition too closely, because when the length scale $\xi$ of
Eq. (\ref{xibulk}) becomes comparable to the system size it is impossible
to separate the edge and the bulk.

Another consequence of the finite system size is that the spectrum is
discrete. So in computing the spectral density of Eq. (\ref{specden})
we replace the Dirac $\delta$-functions by Lorentzians of width
$\eta=0.05$, which produces fairly smooth spectral densities.

Below we use the parameters of the model given in Section \ref{HFsec}. In particular, the
critical point is at $E_Z=0.2$.

\subsection{Bulk Collective Modes}
\label{bulkNMsec}

We begin by presenting the evolution of the bulk collective modes
as $E_Z$ increases. Fig. \ref{bulk-NM-ez0p10fig} shows them deep in the CAF phase
at $E_Z=0.1$. As expected from the spontaneously broken symmetry, the
lowest bulk mode (black line) is a gapless linearly dispersing
Goldstone mode. The next mode (blue) is the Larmor mode, and goes to
the limit $\omega(q=0)=2E_Z$. The highest energy mode is two-fold degenerate.

In Fig. \ref{bulk-NM-ez0p30fig},  we present the bulk
modes at $E_Z=0.3$, deep in the FM phase. There is no
spontaneously broken symmetry, so there is no gapless bulk mode in this
phase. The gap for the lowest mode is $\Delta=2(E_Z-E_{Zc})$. Fig.
\ref{comparisonfig} shows the evolution of the lowest-lying collective bulk
mode as a function of $E_Z$. It is evident that the spin-wave velocity
in the CAF phase vanishes continuously as the transition is
approached. At $E_Z=E_{Zc}$ the lowest-lying mode becomes
quadratically dispersing, and for $E_Z>E_{Zc}$ it ``lifts off'' and
becomes gapped.

Now we are ready to look at the correlators in the bulk. Within the
$\nu=0$ Landau level, in a translationally invariant HF state, the
charge density operator does not couple to leading order to the
collective excitations. We will thus restrict ourselves to the spin
correlators in the bulk.

\subsection{Bulk Spin Correlators}
\label{bulk-spin-corr-subsec}

\begin{figure}[t]
\includegraphics[width=0.8\linewidth]{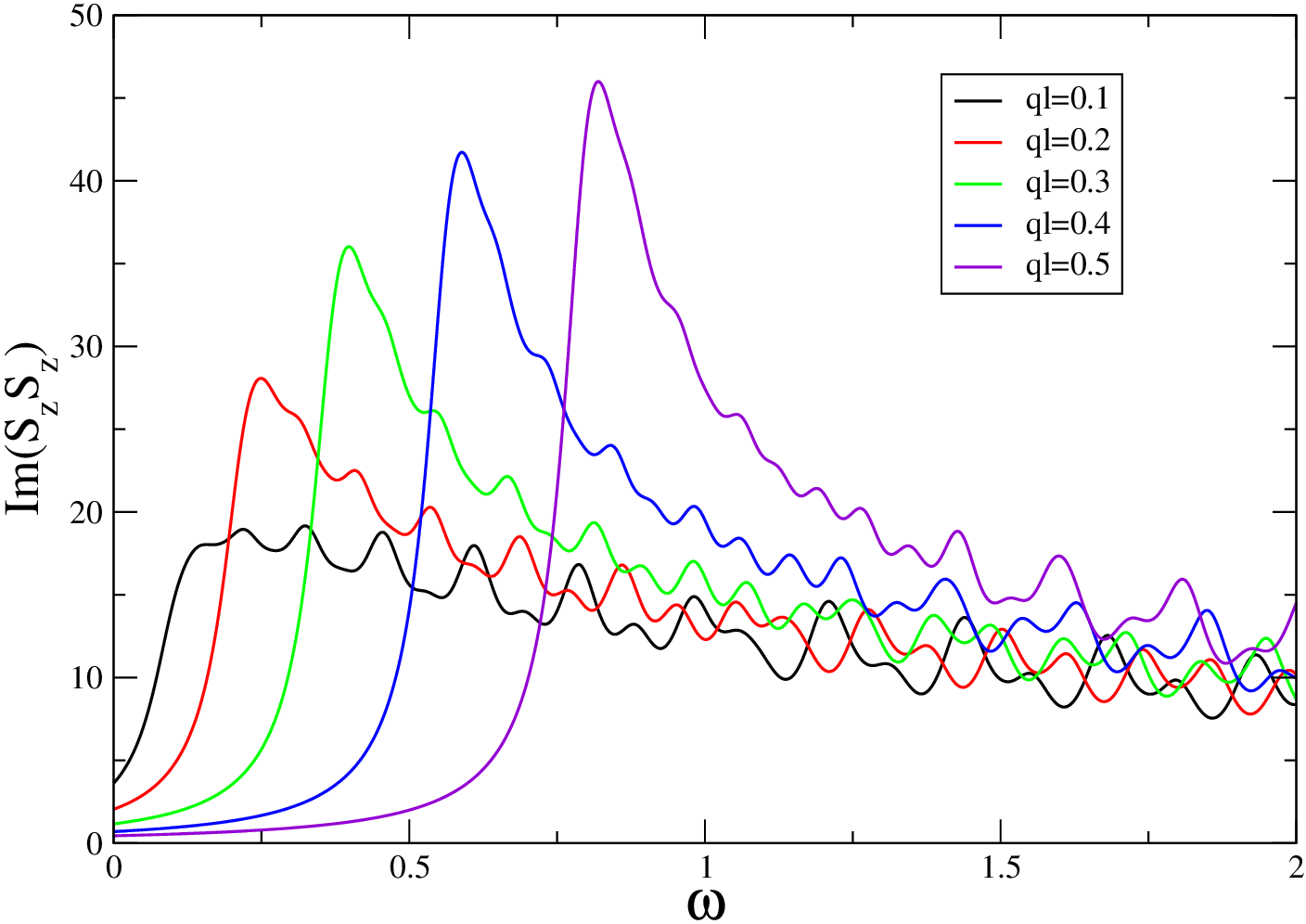}
\caption{(Color online.)  The spectral density of the $S_zS_z$ correlator at
$E_Z=0.1$, deep in the CAF phase. The coupling to the gapless Goldstone mode can be seen.  }
\label{szsz-bulk-ez0p10fig}
\end{figure}

\begin{figure}[h]
\includegraphics[width=0.8\linewidth,trim={0 0 0 0.25cm},clip]{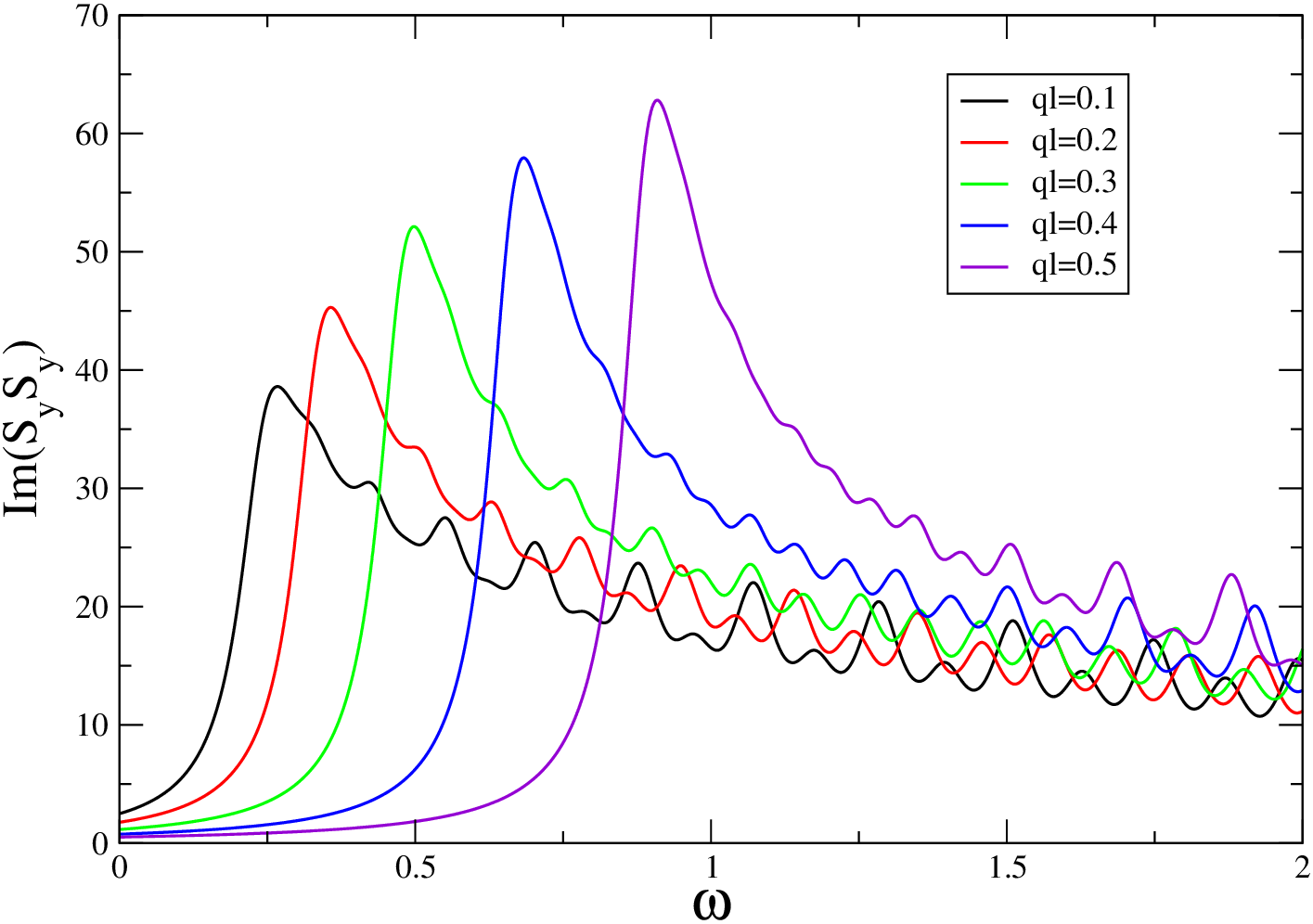}
\caption{(Color online.)  The spectral density of the $S_yS_y$
  correlator at $E_Z=0.1$, deep in the CAF phase. The coupling to the
  gapped Larmor mode can be seen.  }
\label{sysy-bulk-ez0p10fig}
\end{figure}

We begin with the $S_zS_z$ correlator for the CAF phase. In Fig.
\ref{szsz-bulk-ez0p10fig} we show this correlator in the bulk at $E_Z=0.1$, deep in the CAF phase. 
\vskip 0.25in
Due to the condensation of $S_x$, the operator $S_z$ is subject to
quantum fluctuations, and couples strongly to the Goldstone mode. In principle
this is an unambiguous way of detecting the CAF phase.
The $S_xS_x$ and $S_yS_y$ correlators, on the other hand, couple only
to the Larmor mode, and their spectral densities are correspondingly
gapped, as seen in Fig. \ref{sysy-bulk-ez0p10fig}.
\vskip 0.1in
\begin{figure}[t]
\includegraphics[width=0.8\linewidth,trim={0 0 0.25cm 0},clip]{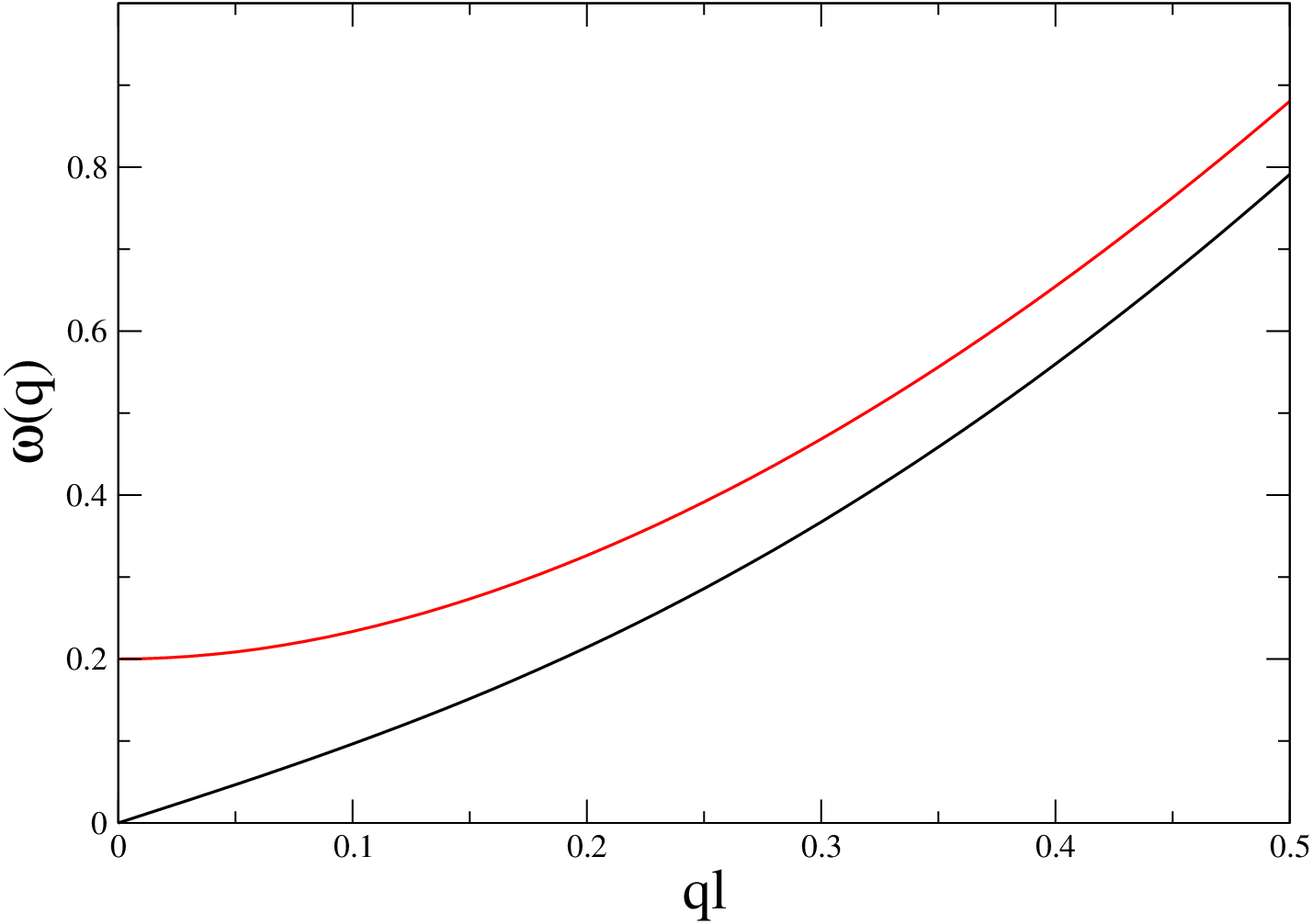}
\caption{(Color online.)  An expanded view of the lowest two modes at $E_Z=0.1$.   }
\label{bulk-NM-ez0p10-expandedfig}
\end{figure}
\vskip 0.25in
To help with the comparison of the peak positions of the spectral
densities, we provide an expanded view of the Goldstone mode and the
Larmor mode for $E_Z=0.1$ in Fig. \ref{bulk-NM-ez0p10-expandedfig}. At
$q_y\ell=0.1$, for example, the $S_zS_z$ spectral function peaks at
$\omega\approx0.1$, which is the Goldstone mode energy, while the
$S_yS_y$ spectral density peaks at $\omega\approx0.25$, which is the
Larmor mode energy. As $q_y$ increases, the difference in peak position persists, but
becomes smaller as the modes become similar in energy.

\begin{figure}[b]
\includegraphics[width=0.8\linewidth,trim={0 0 0 0.25cm},clip]{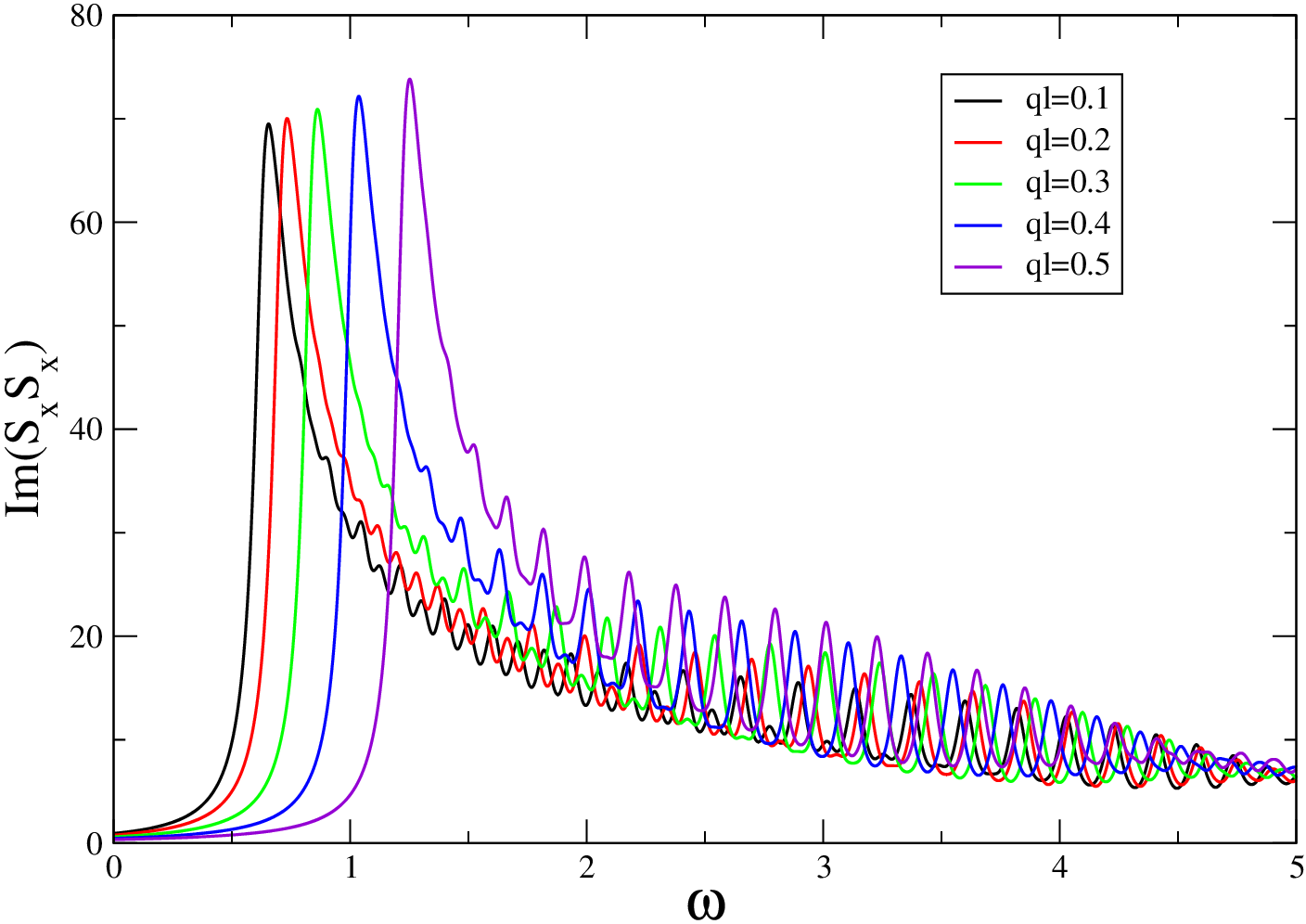}
\caption{(Color online.)  The spectral density of the $S_xS_x$ correlator at $E_Z=0.3$, deep in the FM phase.  }
\label{sxsx-bulk-ez0p30fig}
\end{figure}
\vskip 0.25in

Now we go deep into the FM phase. Here $S_z$ is a good
quantum number, so there are no fluctuations and the $S_zS_z$
correlator is trivial. The $S_xS_x$ and $S_yS_y$ correlators once
again follow the Larmor mode, as shown in Fig. \ref{sxsx-bulk-ez0p30fig}. The gap is larger ($2E_Z=0.6$) and therefore easier to see than at $E_Z=0.1$.

\subsection{Edge Modes and Correlators}
\label{edge-corr-subsec}

\begin{figure}[t]
\includegraphics[width=0.8\linewidth]{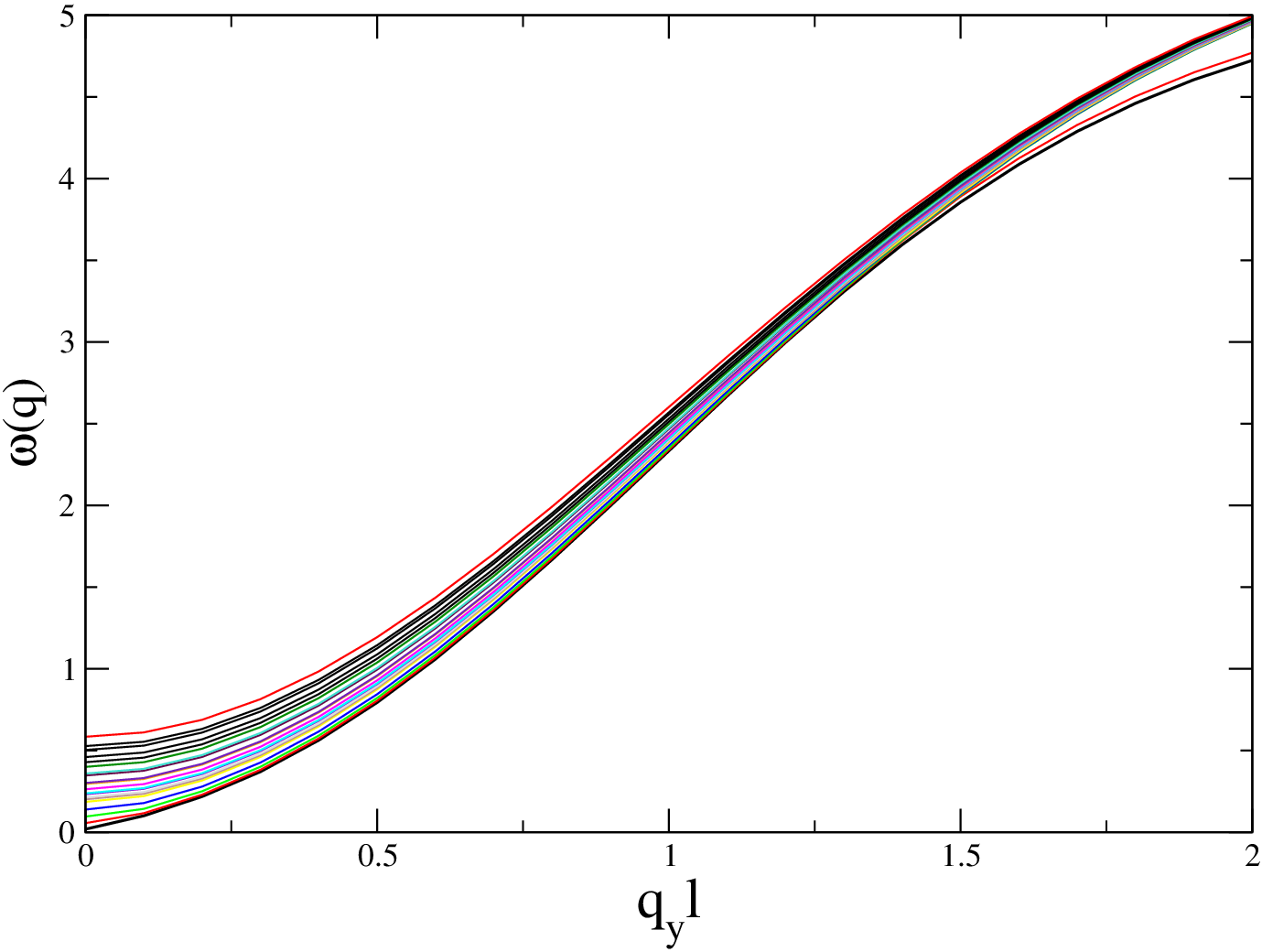}
\caption{(Color online.)  The energy dispersion of collective
  particle-hole excitations vs $q\ell$ for a system with an edge at
  $E_Z=0.1$ in the CAF phase. The figure looks dense because only
  $q_y$ is a good quantum number, so all the possible values of $q_x$
  may get mixed.  }
\label{edge-modes-ez0p10fig}
\end{figure}

Let us start with the dispersion of collective particle-hole modes in
a system with an edge. Fig. \ref{edge-modes-ez0p10fig} shows the first
few modes at $E_Z=0.1$, in the CAF phase. Since only $q_y$ is a good
quantum number, all the values of $q_x$, which were good quantum
numbers in the bulk, are now potentially mixed. The bottom of the
quasi-continuum is the bulk gapless mode, shown in the bold black
line.

\vskip 0.25in

\begin{figure}[htb]
\begin{center}
\includegraphics[width=0.8\linewidth,trim={0 0 0.25cm 0},clip]{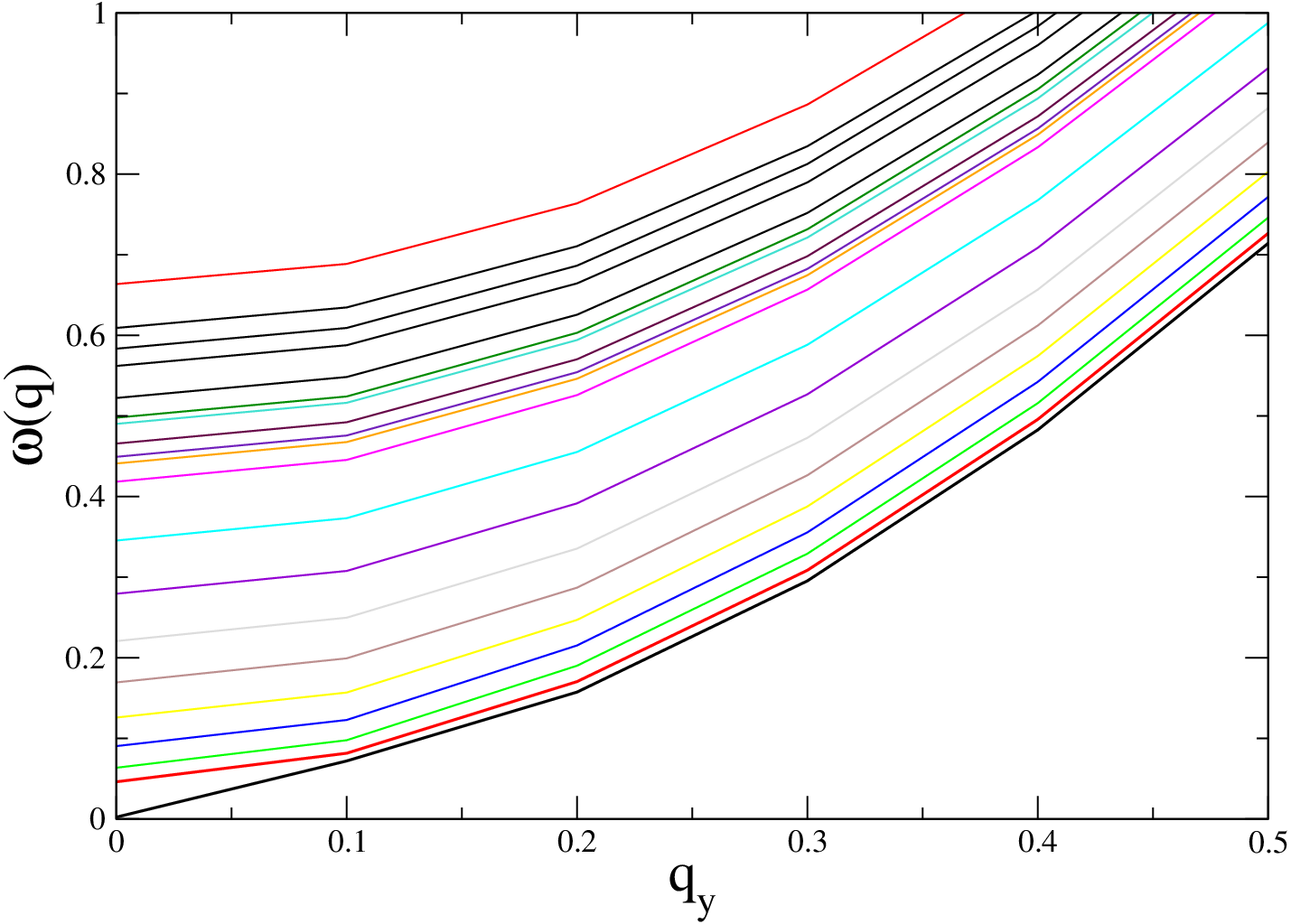}
\caption{(Color online.)  The energy dispersion of collective
  particle-hole excitations vs $q\ell$ for a system with an edge at
  $E_Z=0.22$, close to the transition in the FM phase. Note the gapless mode (thick black line) which was absent in the bulk spectrum of Fig. \ref{bulk-NM-ez0p30fig}.  }
\label{edge-modes-ez0p22fig}
\end{center}
\end{figure}

\begin{figure}[t]
\includegraphics[width=0.8\linewidth]{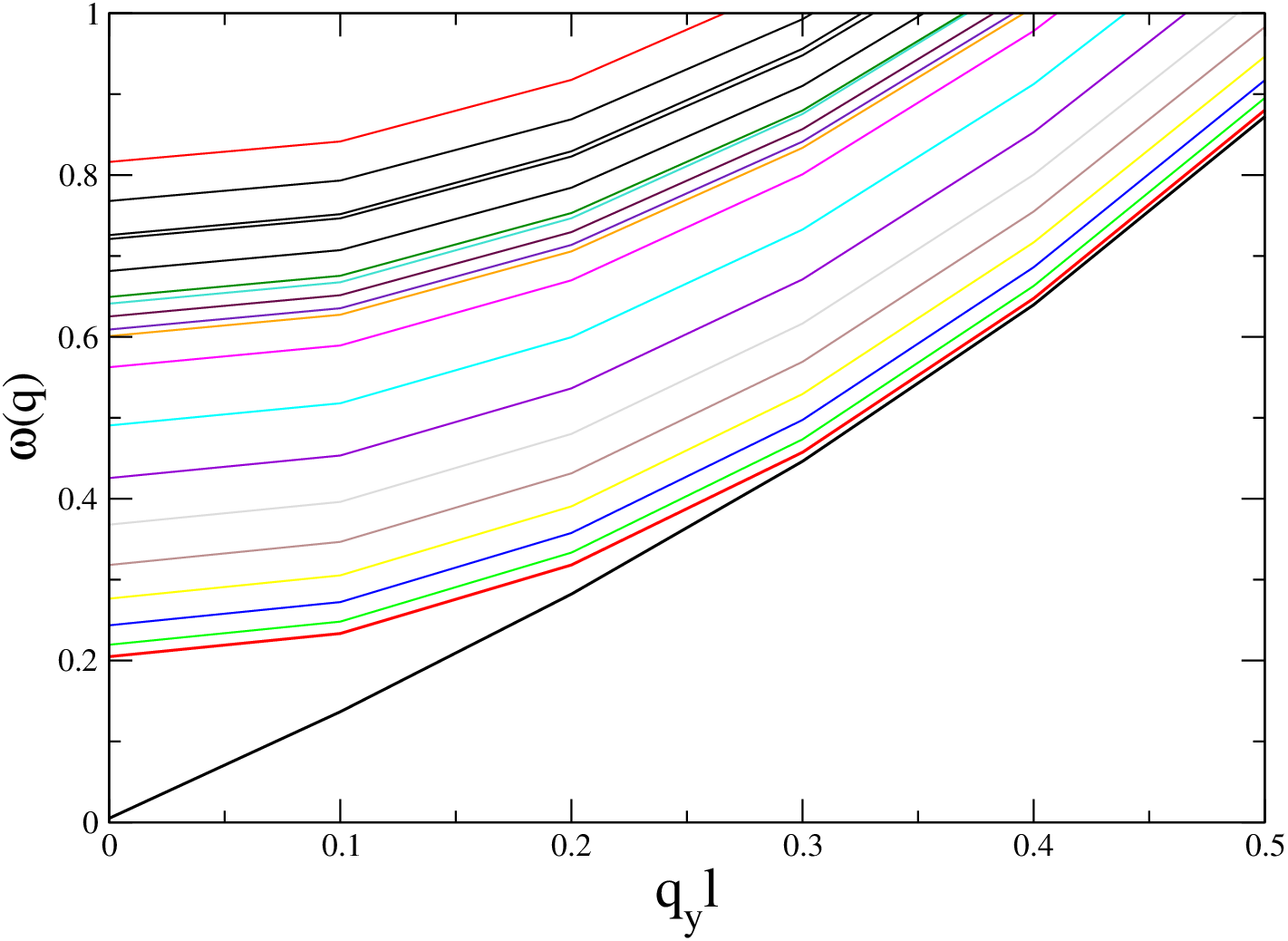}
\caption{(Color online.)  The energy dispersion of collective
  particle-hole excitations vs $q\ell$ for a system with an edge at
  $E_Z=0.3$, deep in the FM phase. Note the gapless mode
  (thick black line) which was absent in the bulk spectrum of
  Fig. \ref{bulk-NM-ez0p30fig}.  }
\label{edge-modes-ez0p30fig}
\end{figure}

Things become more interesting when we go to the FM phase. Recall
that, as seen in Fig. \ref{bulk-NM-ez0p30fig}, the bulk was gapped in
this phase. The first few modes in the system with an edge at
$E_Z=0.22$ are shown in Fig. \ref{edge-modes-ez0p22fig}. As can be
seen there is now a gapless mode (thick black line) which was not
present in the bulk system. This becomes even clearer when one goes
deeper into the FM phase, as shown in
Fig. \ref{edge-modes-ez0p30fig}. Thus the TDHFA supports the
expectation that the FM state, despite having a gapped HF spectrum,
supports gapless edge modes \cite{Kharitonov_edge,Murthy2014}. This is
another example of the way TDHF restores the symmetry broken by the HF
approximation. The naive view, that the gapless mode is the Goldstone
mode of the symmetry broken in HF, is incorrect in this case: In a 1D
system, a continuous symmetry cannot be broken even at $T=0$, and the
symmetry-breaking seen in HF is an artefact.

\vskip 2.5in

\vskip 0.25in

\begin{figure}[b]
\includegraphics[width=0.8\linewidth,trim={0 0 0 0.25cm},clip]{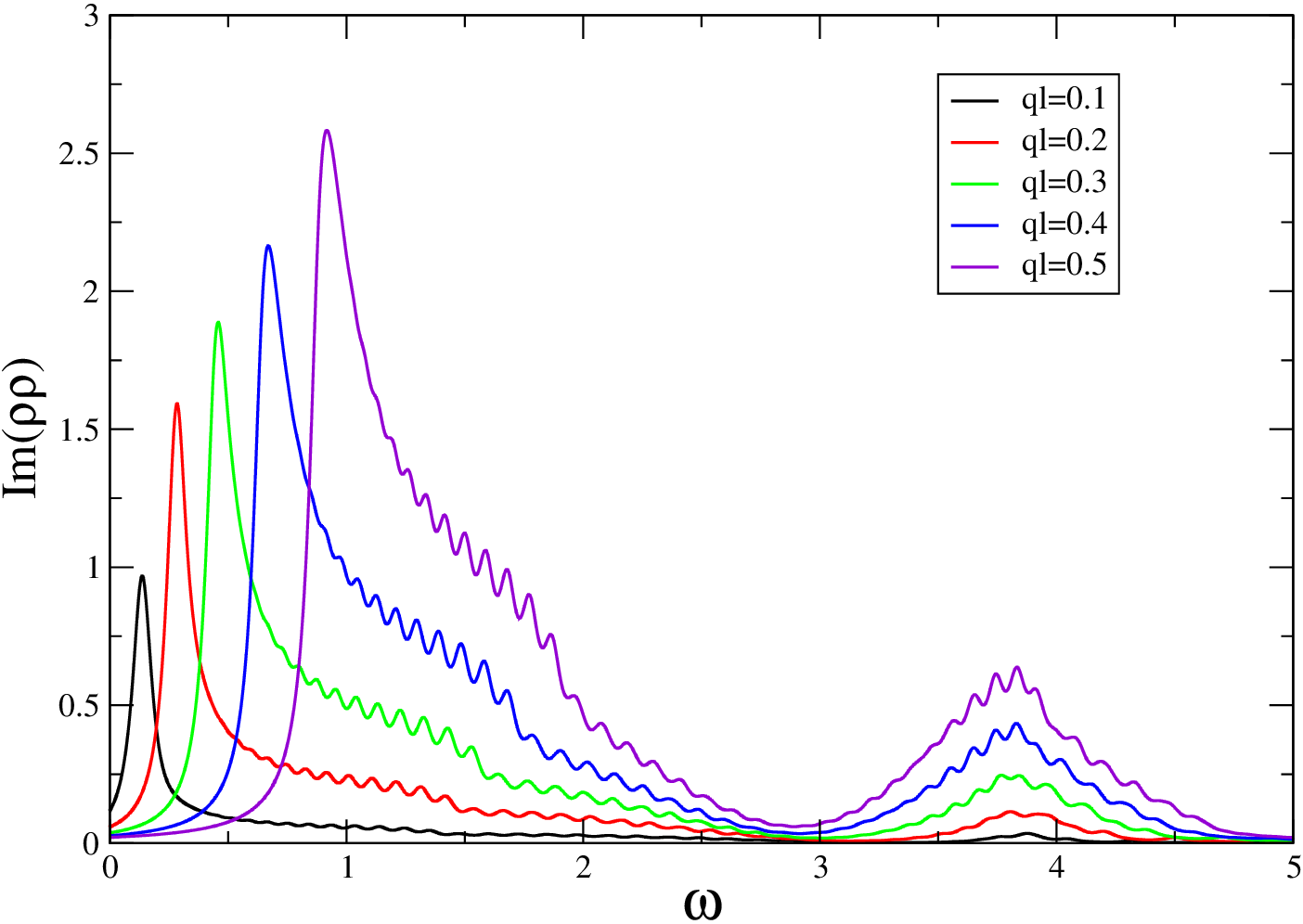}
\caption{(Color online.)  The spectral density of the charge
  density-density correlator as a function of $\omega$ for different
  values of $q_y\ell$ at $E_Z=0.3$ deep in the FM
  phase. Note the peaks at low $\omega$ which correspond to the
  gapless edge modes in Fig. \ref{edge-modes-ez0p30fig}.  }
\label{rhorho-ez0p30fig}
\end{figure}

\begin{figure}[t]
\begin{center}
\includegraphics[width=0.8\linewidth,trim={0 0 0.15cm 0},clip]{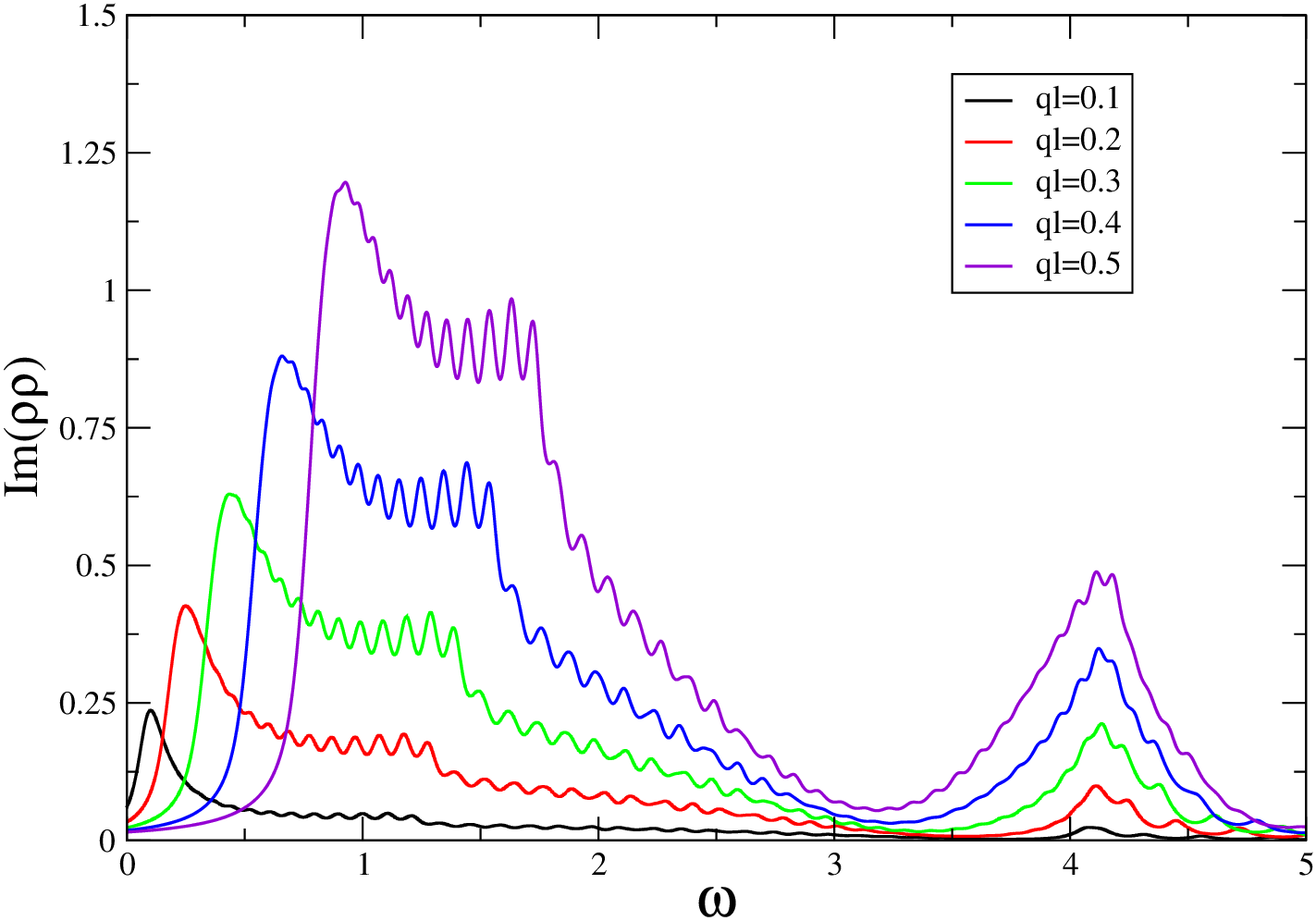}
\caption{(Color online.)  The spectral density of the charge
  density-density correlator as a function of $\omega$ for different
  values of $q_y\ell$ at $E_Z=0.22$, close to the transition but in
  the FM phase. Note the peaks at low $\omega$ which
  correspond to the gapless edge modes in
  Fig. \ref{edge-modes-ez0p22fig}.}
  \label{rhorho-ez0p22fig}
\end{center}
\end{figure}

There is another important aspect to the physics of the gapless edge
mode: it must be able to carry charge. To ascertain that this is
indeed true we look at the spectral density of the charge-charge
correlator in the system with an edge. Fig. \ref{rhorho-ez0p30fig}
shows the spectral density of the charge correlator at $E_Z=0.3$, deep
in the FM phase. The peaks dispersing linearly as a
function of $q\ell$ show that the gapless edge mode indeed carries
charge. Fig. \ref{rhorho-ez0p22fig} shows that this persists close to
the transition. However, in this situation, the gapless edge
mode admixes with low-energy gapped bulk modes, leading to some
broadening. (This may have important consequences for transport at
finite temperature, a subject we will address in a future publication.)

We also note that in addition to the gapless edge mode, the charge
density correlator also couples to a high-energy mode (with an energy
around $\omega\approx4$ in our units). This could be a gapped
charge-carrying mode bound to the edge, and seems to stiffen as one
approaches the critical point.

\vskip 2.5in

\vskip 0.25in

\begin{figure}[b]
\includegraphics[width=0.8\linewidth]{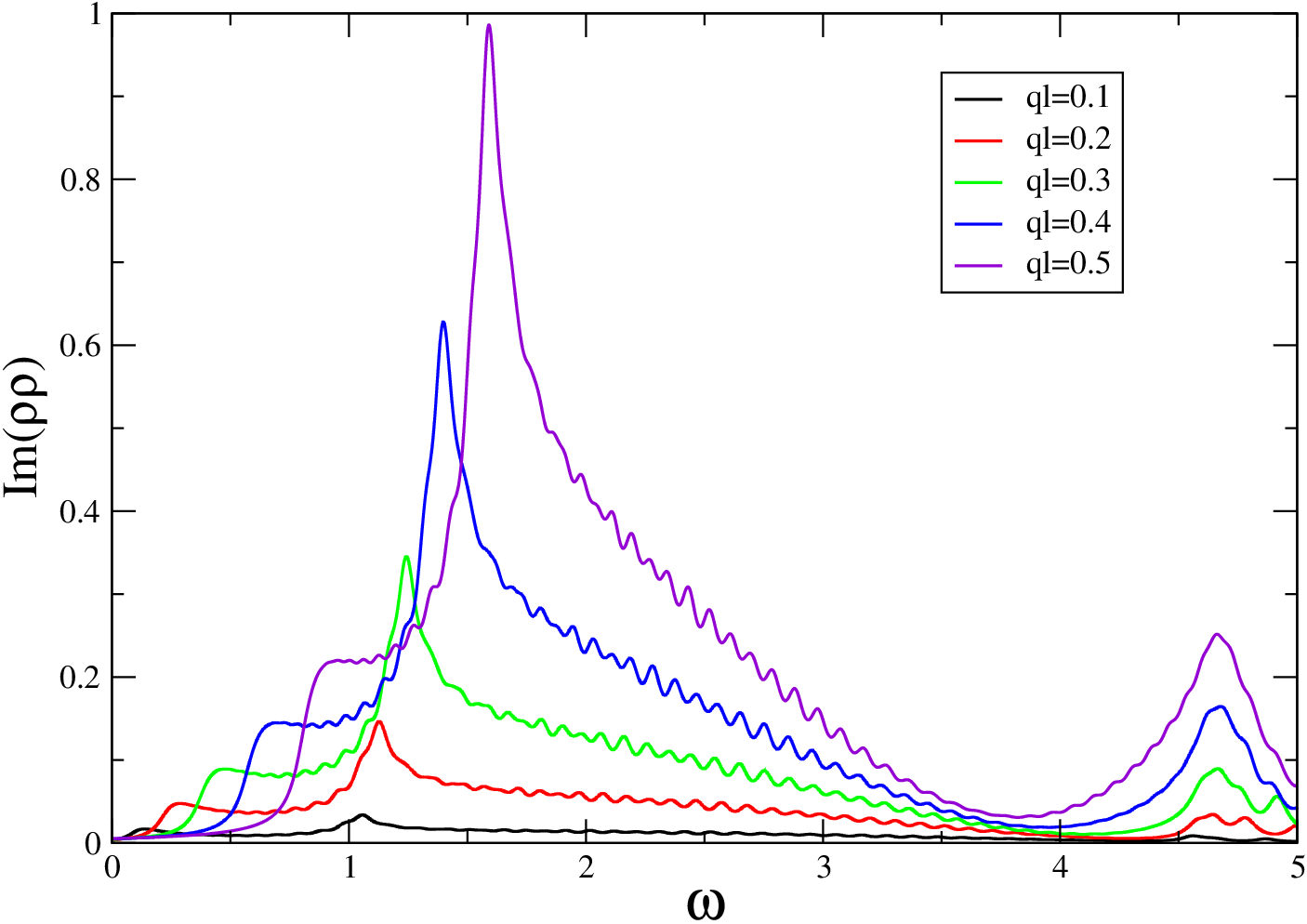}
\caption{(Color online.)  The spectral density of the charge
  density-density correlator as a function of $\omega$ for different
  values of $q_y\ell$ at $E_Z=0.10$, deep in the CAF phase. Note that
  the peaks in the spectral density are gapped, indicating that the
  charge does not couple to the gapless Goldstone mode.}
\label{rhorho-ez0p10fig}
\end{figure}

\begin{figure}[t]
\includegraphics[width=0.8\linewidth]{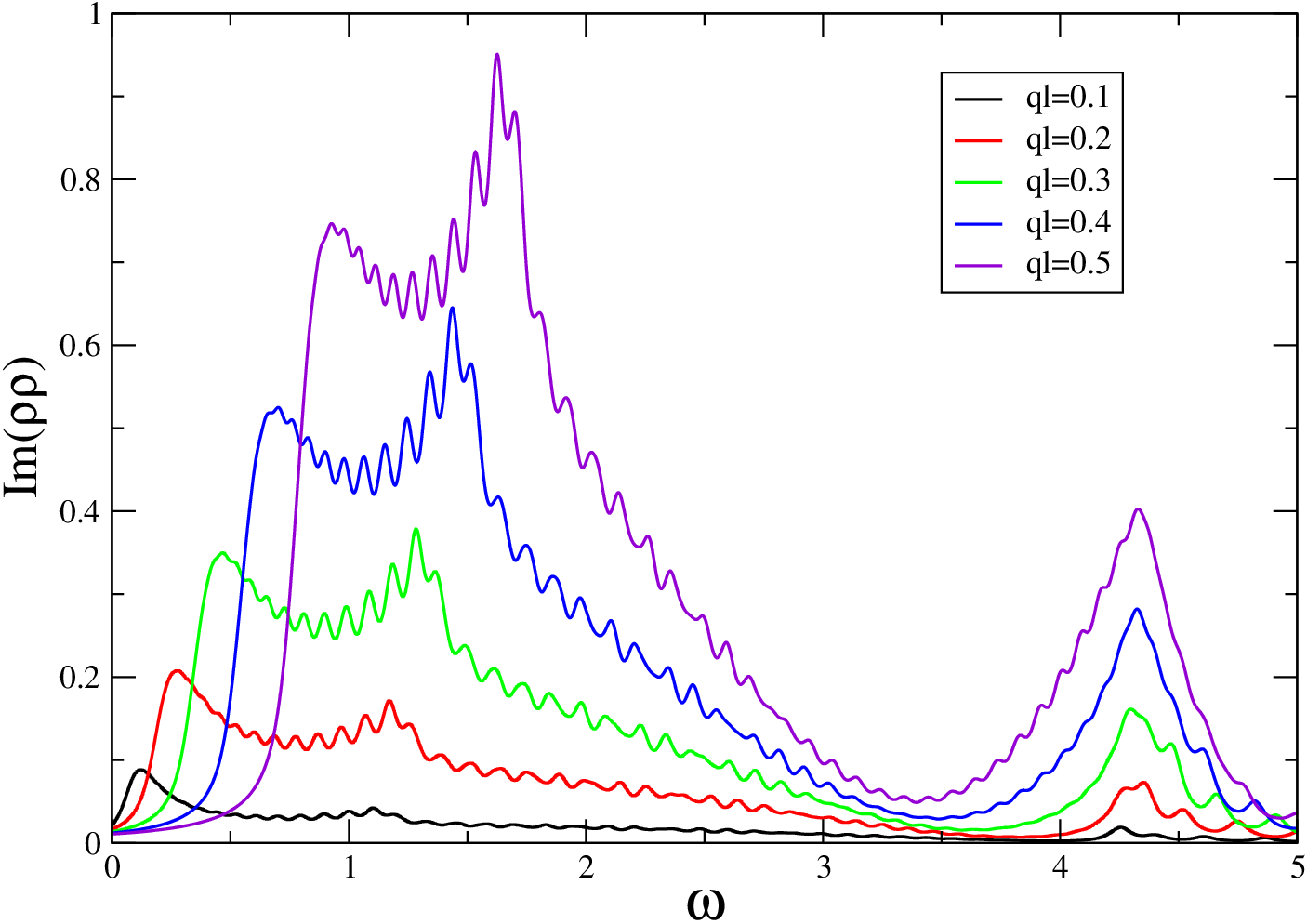}
\caption{(Color online.)  The spectral density of the charge
  density-density correlator as a function of $\omega$ for different
  values of $q_y\ell$ at $E_Z=0.18$, close to the transition in the
  CAF phase. There are now contributions from both gapless and gapped
  modes, presumably due to finite-size effects. }
\label{rhorho-ez0p18fig}
\end{figure}

In Fig. \ref{rhorho-ez0p10fig} we show the spectral density of the charge
correlator at $E_Z=0.1$, deep in the CAF phase. The gapped nature of
the excitations coupling to charge is evident. As one approaches very
close to the transition, the finite size effects mentioned at the
beginning of the section come into play. The quantum phase transition,
which would have been sharp in a thermodynamically large system,
becomes instead a crossover. This is seen in the spectral density of
charge correlator at $E_Z=0.18$, shown in Fig. \ref{rhorho-ez0p18fig}. As
at $E_Z=0.22$ [Fig. \ref{rhorho-ez0p22fig}], one can see that both the gapless (bulk) mode and a
gapped mode contribute. We have checked that the contribution of the
gapless mode decreases as the system size is increased, whereas the
contribution of the gapped mode does not change. The contribution of the gapped charge-carrying mode noted in the F phase persists in the CAF phase as well. 

\vskip 2.5in

\vskip 0.25in
To complete the picture, let us examine the spin correlators.
\begin{figure}[b]
\includegraphics[width=0.8\linewidth]{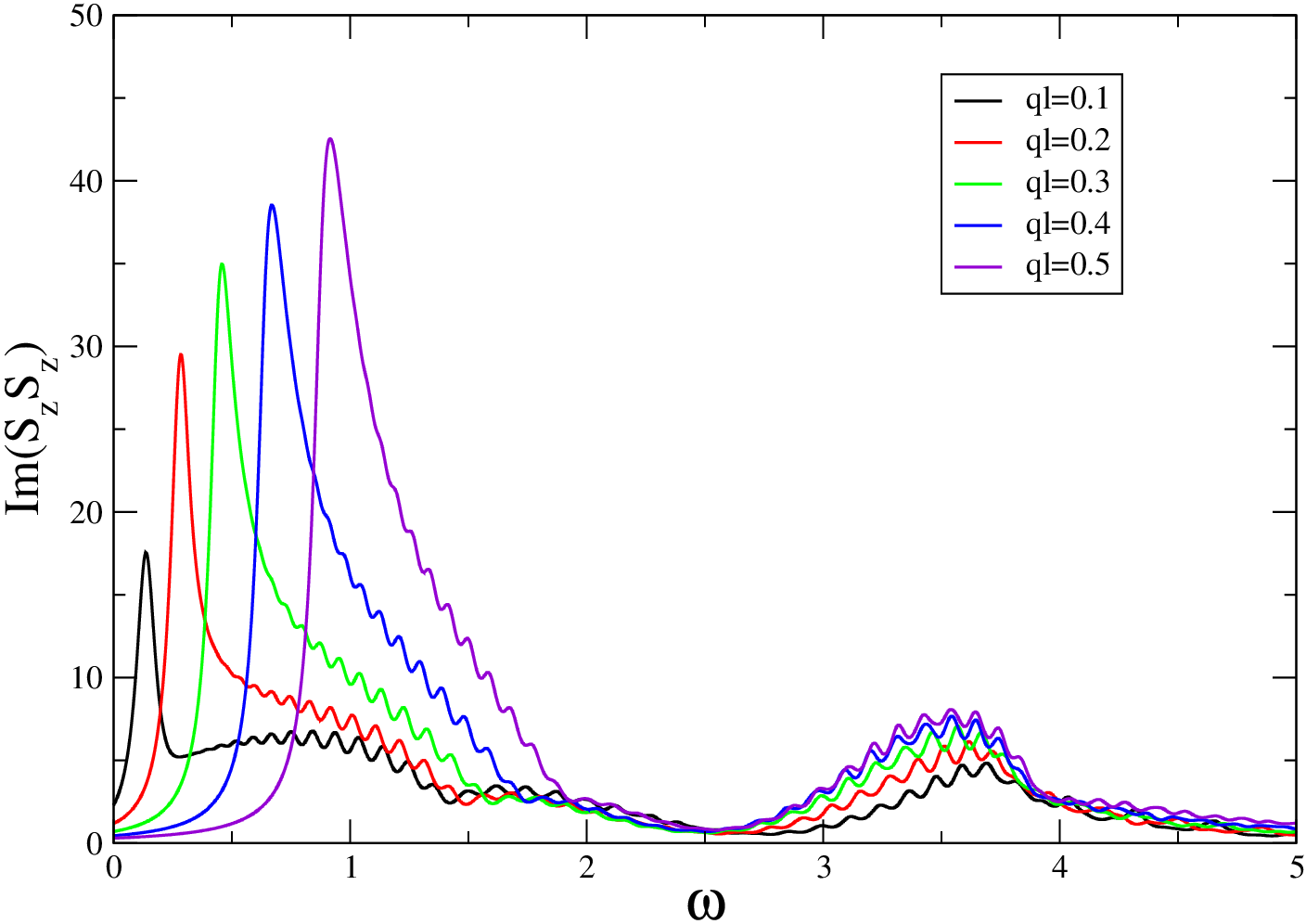}
\caption{(Color online.)  The spectral density of the $S_zS_z$
  correlator as a function of $\omega$ for different values of
  $q_y\ell$ at $E_Z=0.30$, deep in the FM phase. In contrast to the
  bulk correlator, this spectral density couples strongly to the
  gapless edge mode.  }
\label{szsz-edge-ez0p30fig}
\end{figure}
\vskip 0.25in
This time we will start in the FM phase. As noted in the
previous subsection, the bulk $S_zS_z$ correlator is trivial in the
FM phase, because the bulk is fully polarized. In
Fig. \ref{szsz-edge-ez0p30fig} we see that this is not the case when
an edge is present. The spectral density of this correlator couples to
the gapless mode as well. This can be understood from an effective
field theory as follows: The one-dimensional field theory describing
the edge deep in the FM phase is a helical Luttinger liquid \cite{Fertig2006,SFP}, in which
the right-movers have spin up (say) and left-movers have spin down. In
such a system the charge current is proportional to the
$S_z$-density. Thus, it is natural that the $S_zS_z$ correlator
couples to these gapless excitations.
Unfortunately, this by itself cannot be used as a signature of the phase transition
because the qualitative behavior is the same in the CAF phase, as
shown in Fig. \ref{szsz-edge-ez0p10fig}. Here the gapless mode the
correlator couples to is the bulk Goldstone mode.

\begin{figure}[t]
\begin{center}
\includegraphics[width=0.9\linewidth,trim={0 0 0 0.15cm},clip]{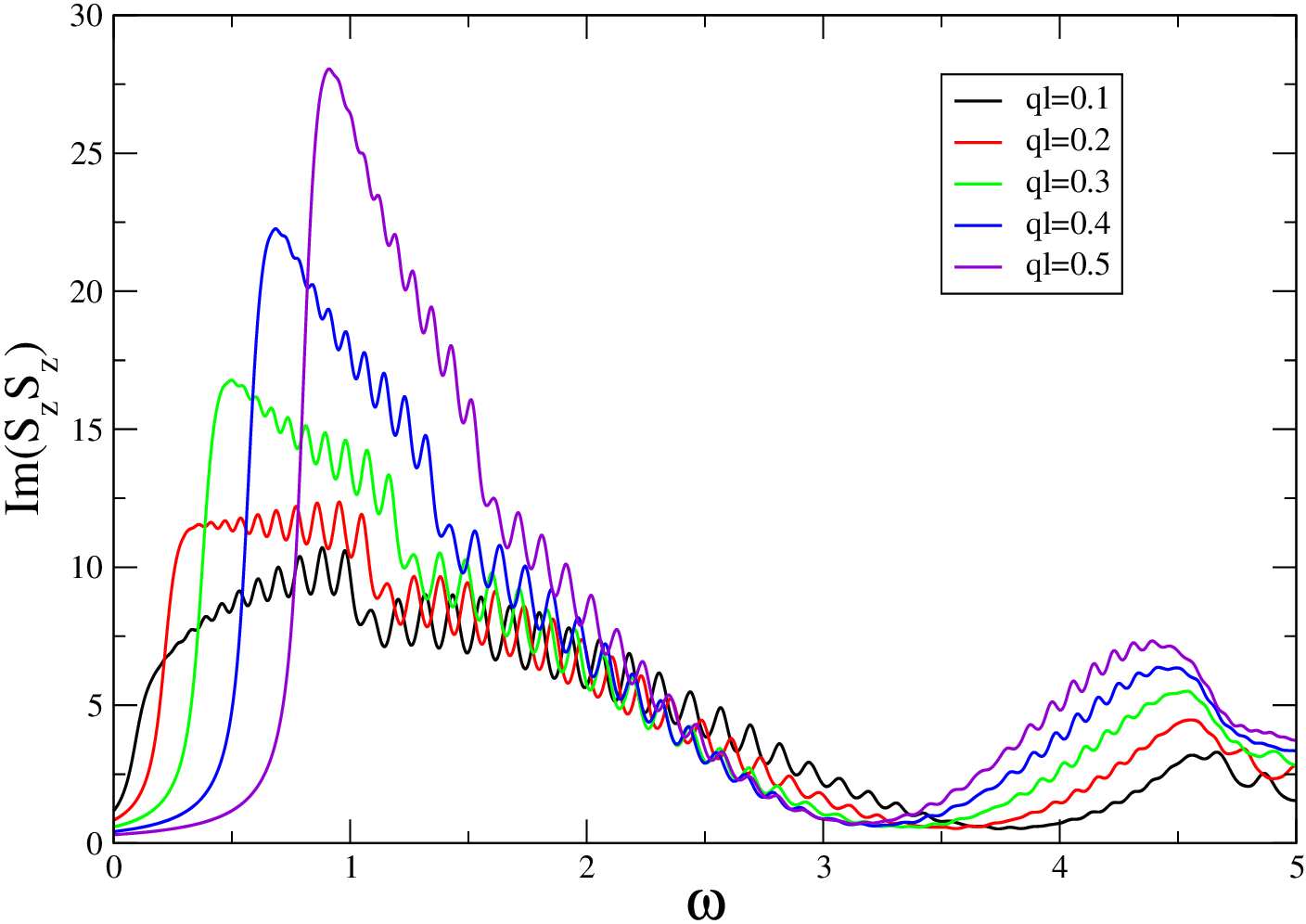}
\caption{(Color online.)  The spectral density of the $S_zS_z$
  correlator as a function of $\omega$ for different values of
  $q_y\ell$ at $E_Z=0.10$, deep in the CAF phase. The peaks now
  indicate a coupling to the Goldstone mode.  }
\label{szsz-edge-ez0p10fig}
\end{center}
\end{figure}
\vskip 0.25in

\subsection{Space and Time-Dependent Response at the Edge}
\label{spacetimesec}

\begin{figure}[b]
\includegraphics[width=0.9\linewidth,trim={0 0 0 0.15cm},clip]{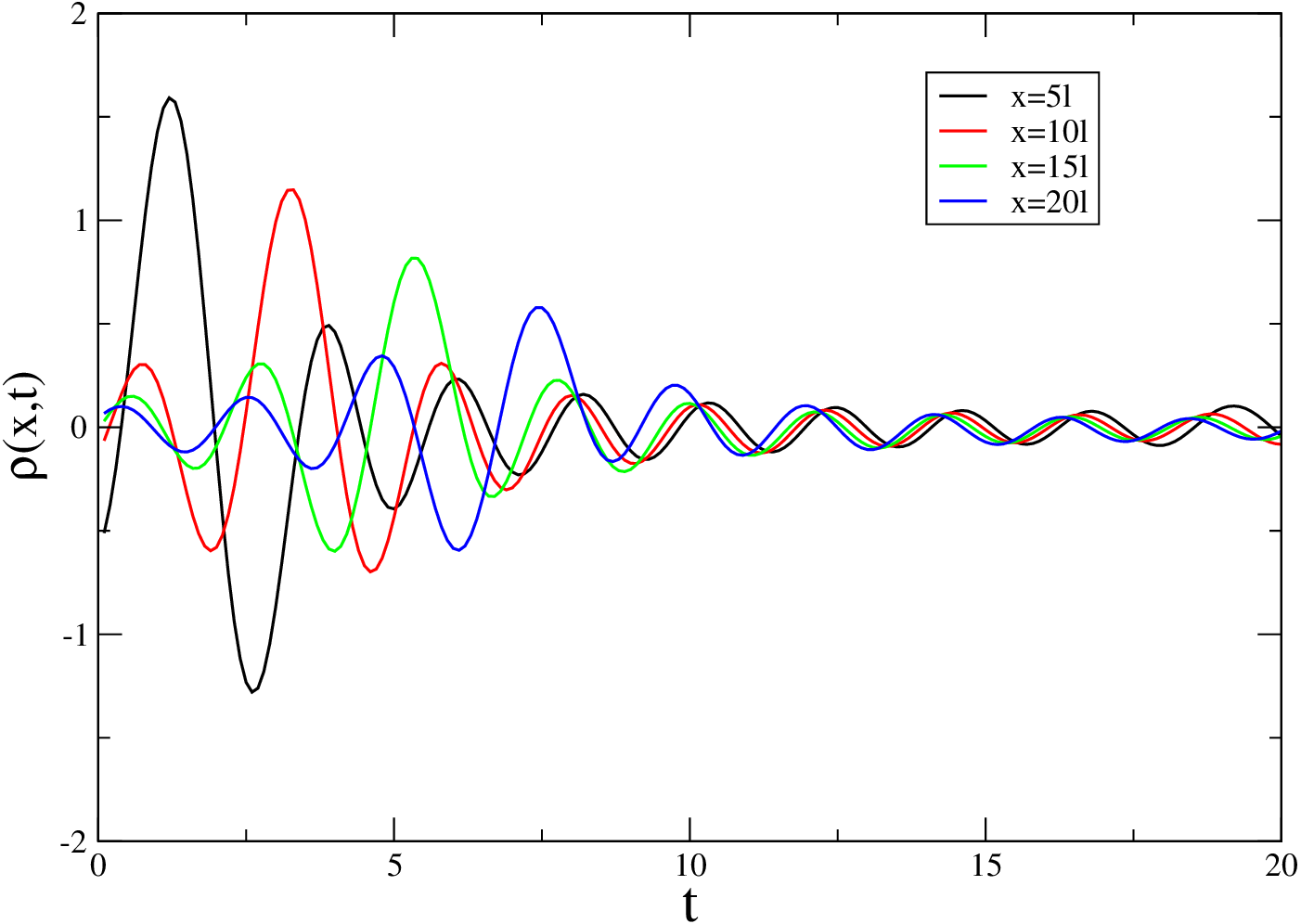}
\caption{(Color online.)  The $t$-dependent response at different
  locations to a localized density perturbation along the edge at
  $E_Z=0.3$. The linear edge mode produces a travelling pulse. }
\label{spacetime-d-ez0p30fig}
\end{figure}

\begin{figure}[t]
\includegraphics[width=0.9\linewidth]{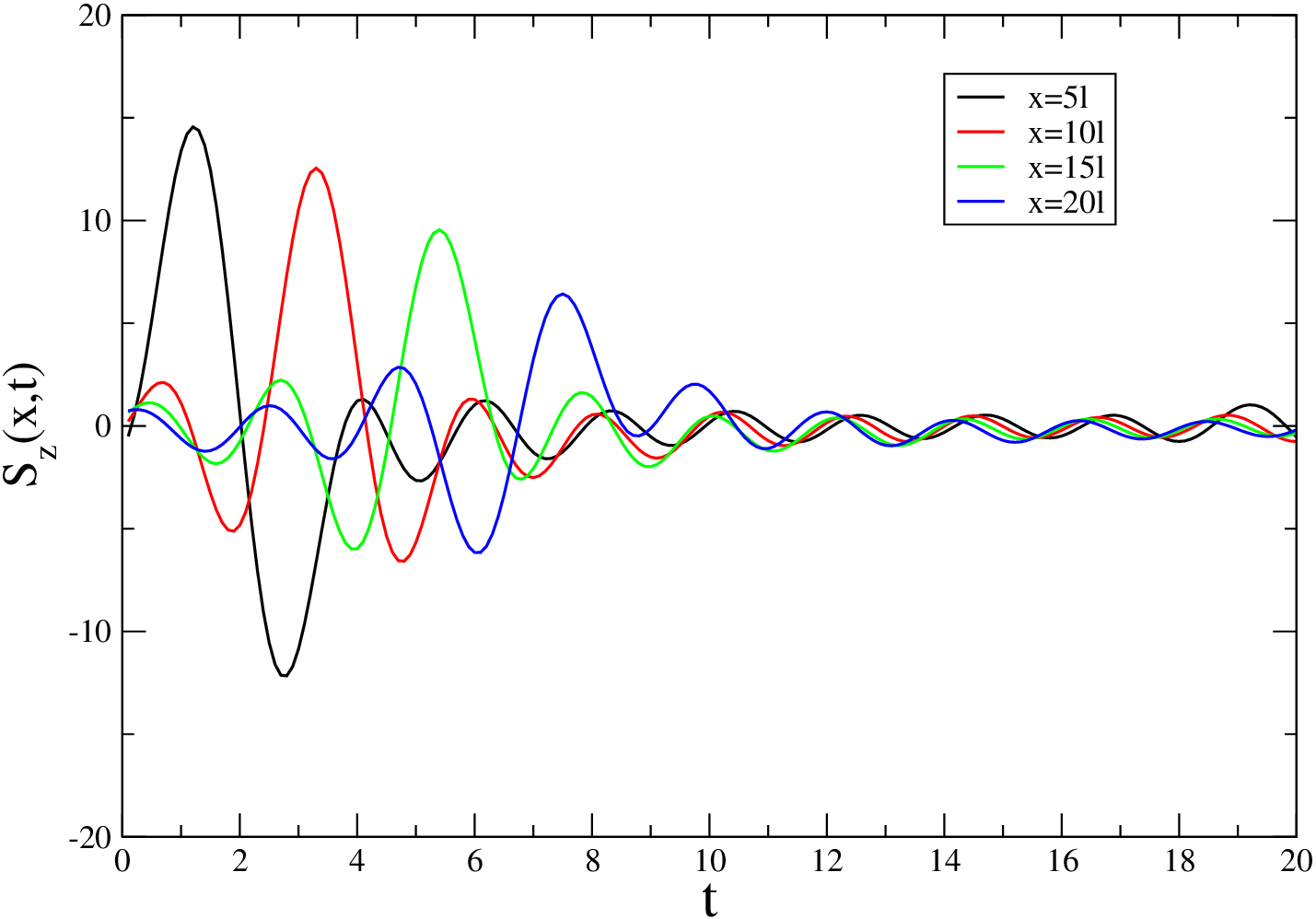}
\caption{(Color online.)  The $t$-dependent response at different
  locations to a localized $S_z$ perturbation along the edge at
  $E_Z=0.3$. The linear edge mode produces a travelling pulse.  }
\label{spacetime-z-ez0p30fig}
\end{figure}

The linearly dispersing mode at the edge can be seen in a much more
physical way. Imagine that we make a localized (in both space and
time) perturbation at a particular position at the edge. If there is a
linearly dispersing mode that couples to the physical perturbation in
question, the effects can propagate arbitrarily far. To be specific,
let us consider a perturbation (induced by, e.g., a field pulse) of the form
\beq
\mH\to \mH+C e^{-\frac{y^2}{2\lambda^2}-\frac{t^2}{2\tau^2}} \mQ(y,t).
\eeq
By expanding $\mQ$ in terms of the eigenoperators of the TDHF
Hamiltonian (Eq. \ref{expandQinb}), after a few straightforward
manipulations we obtain
\beq
\langle \mQ(y,t)\rangle \propto \int\limits_{0}^{\infty}
\frac{d\omega}{\pi} \sum\limits_{\bq}  e^{-(q\lambda)^2/2-t^2/2\tau^2} \sin(qy-\omega t) S_{QQ}(q,\omega).
\eeq
Fig. \ref{spacetime-d-ez0p30fig} illustrates the response to a density
perturbation at the edge (localized at $y=t=0$) deep in the
FM phase measured at different values of $y$ as a function
of time. The propagating mode manifests itself as a peak that shifts
to later times as one moves further away.
%
The same is seen when the perturbation is in $S_z$ instead of $\rho$
(see Fig. \ref{spacetime-z-ez0p30fig}), which is consistent with the interpretation of the edge as a helical Luttinger liquid.

\begin{figure}[b]
\includegraphics[width=0.9\linewidth]{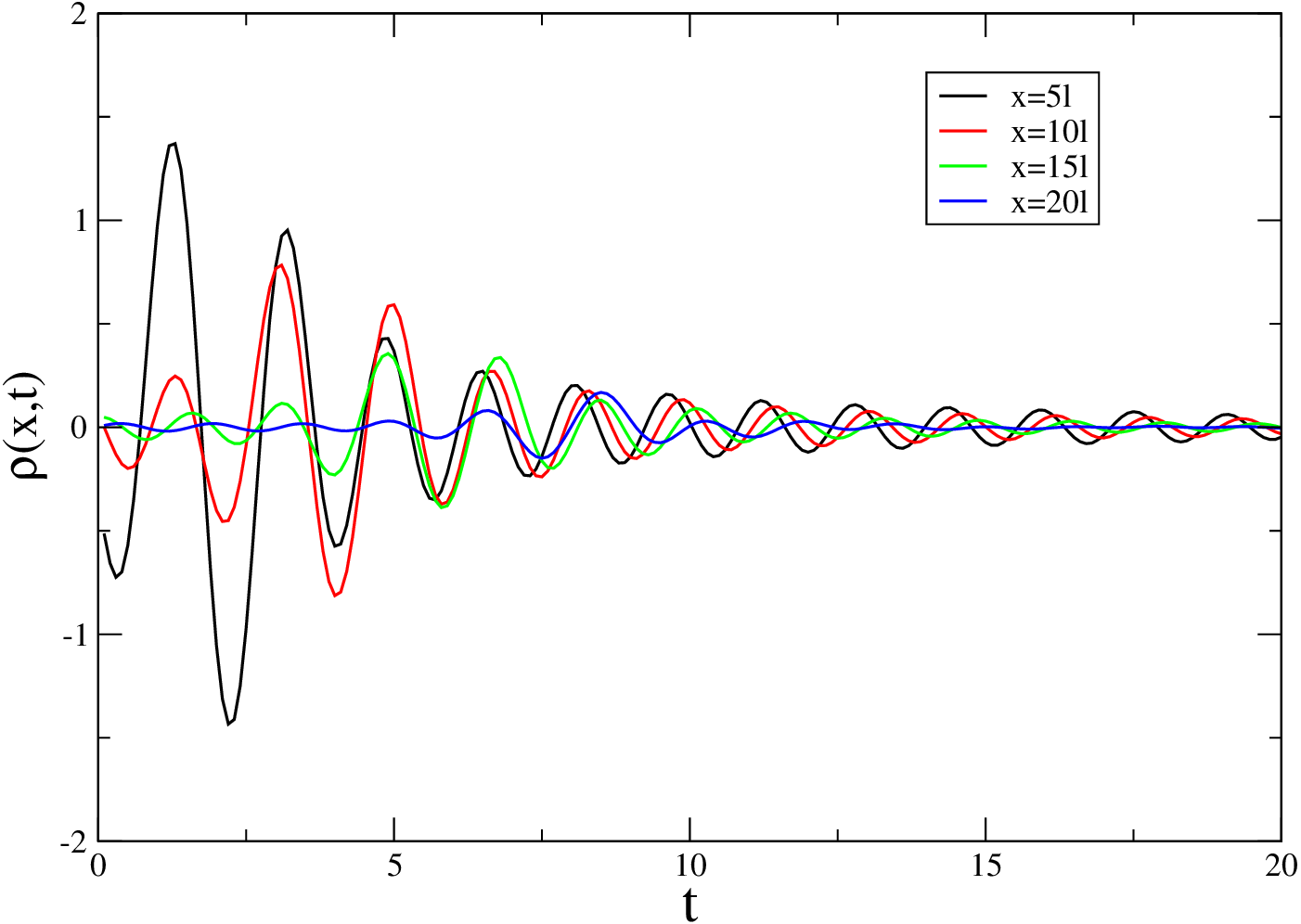}
\caption{(Color online.)  The $t$-dependent response at different
  locations to a localized density perturbation along the edge at
  $E_Z=0.1$. There is no travelling pulse showing the absence of an
  edge mode coupling to charge.  }
\label{spacetime-d-ez0p10fig}
\end{figure}

\begin{figure}[t]
\begin{center}
\includegraphics[width=0.9\linewidth,trim={0 0 0.15cm 0},clip]{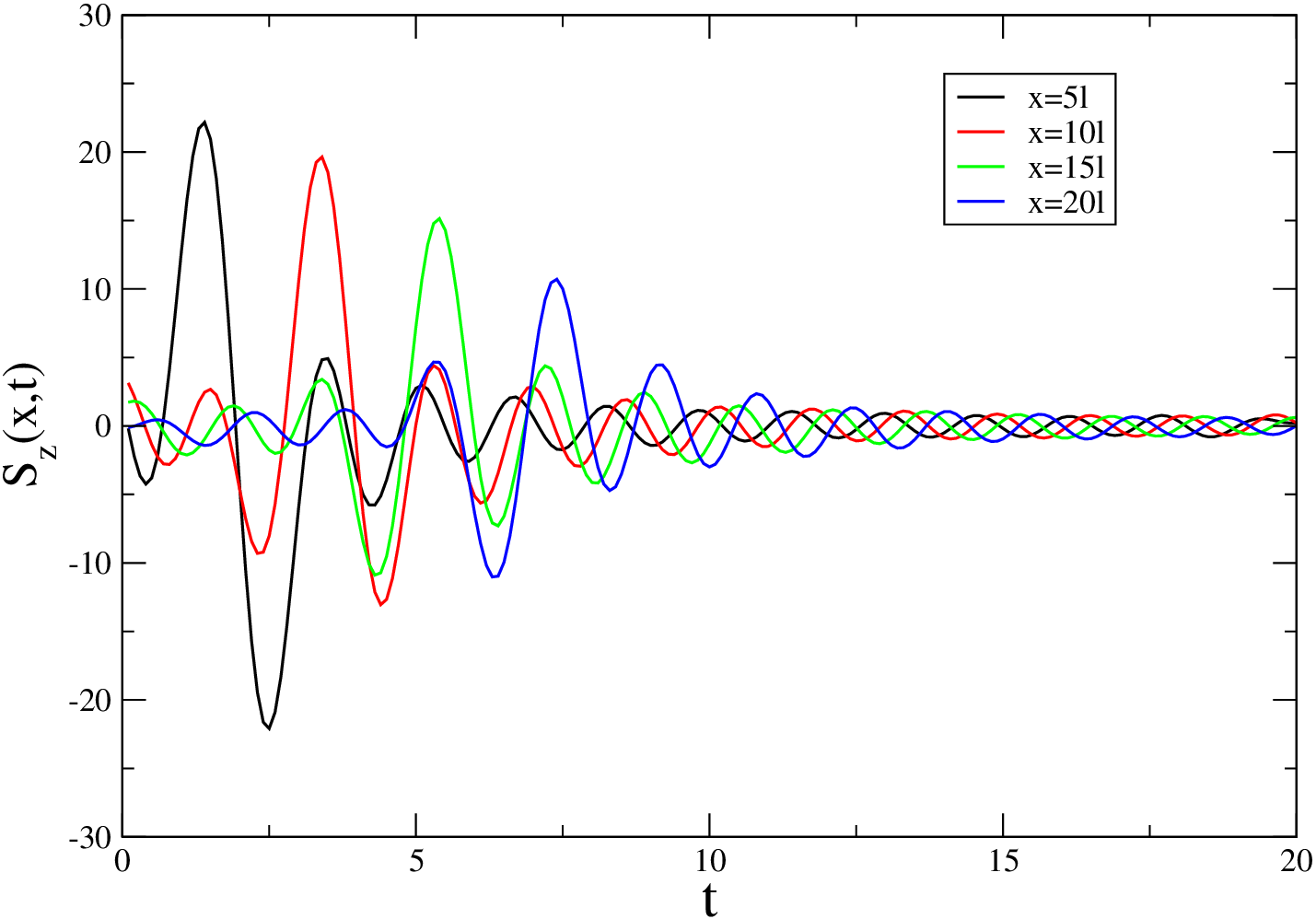}
\caption{(Color online.)  The $t$-dependent response at different
  locations to a localized $S_z$ perturbation along the edge at
  $E_Z=0.1$. The coupling to the bulk Goldstone mode produces a
  travelling pulse.  }
\label{spacetime-z-ez0p10fig}
\end{center}
\end{figure}

When we go deep into the CAF phase, we do not expect a propagating
edge mode that couples to density. As seen in
Fig. \ref{spacetime-d-ez0p10fig} the response as a function of time is
only weakly dependent on the position. However, if the perturbation is
in $S_z$, Fig. \ref{spacetime-z-ez0p10fig} shows that there is a
propagating mode, which we can assume to be the bulk Goldstone mode.

\vskip 2.5in

\vskip 0.25in

\section{Summary, Conclusions and Open Questions}
\label{conclusionsec}

In this paper we investigate the nature of collective particle-hole
excitations in $\nu=0$ single-layer graphene. This system has been
shown experimentally\cite{Young2013} to undergo a quantum phase
transition as a function of Zeeman coupling $E_Z$. For $E_Z<E_{Zc}$
the state is an insulator, while for $E_Z>E_{Zc}$ there are conducting
edge modes robust to disorder. A simple model proposed by
Kharitonov\cite{Kharitonov_bulk,Kharitonov_edge} displays precisely
such a phase transition, explaining it as a transition from a canted
antiferromagnet (CAF) phase in which charge modes are fully gapped to
a fully polarized ferromagnetic (FM) phase which has gapless edge
modes.

In previous work \cite{Murthy2014} we carried out a static
Hartree-Fock analysis on Kharitonov's model in a system with an edge,
showing that the occupied manifold could be characterized by two
angles $\psi_{a,b}$ which characterized entanglement between the spin
and valley sectors. These angles became equal deep in the bulk, but
differed near the edge. We proposed an ansatz for charge excitations
bound to the edge, and showed that while in the CAF phase they are
gapped, they become gapless in the FM phase.

In this paper these ideas are substantiated in the time-dependent
Hartree-Fock (TDHF) approximation
and
physically measurable correlation functions are computed, both for the bulk and
the edge.

In the bulk FM phase, the density and $S_z$ correlators are
fully gapped. As one goes through the transition into the CAF phase
there is a divergent length scale, associated with the vanishing of
the gap of the critical mode. At the critical point it becomes
quadratically dispersing.

Unfortunately, there seems to be no simple way to probe the critical
mode in the bulk FM phase. It does not couple to any of the
natural physical observables, such as components of spin or the charge
density. It may, in principle, be possible to infer its existence by
indirect means. For example, when the mode gets low enough, it should
hybridize with sound waves, and may show up in acoustic
attenuation. If some analog of inelastic light scattering were
possible in single-layer graphene, it should be visible there as well.

In the bulk CAF phase the symmetry breaking represented by the angles
$\psi_{a,b}\ne0$ leads to a neutral Goldstone mode, which can be seen
in the $S_zS_z$ correlator. The $S_xS_x$ and $S_yS_y$ correlators have
spectral densities coupling to the Larmor mode, which is gapped in
both phases and through the transition.  The signature of the bulk CAF
phase is the gaplessness of the spectral density of the $S_zS_z$
correlator. This spectral density becomes gapped at the phase
transition. In principle, the gapless $S_zS_z$ correlator could be
used to distinguish the CAF phase from other proposals for the $\nu=0$
QH state, such as singlet Kekule\cite{Kekule} or charge density wave phases.

Coming now to the edge, we clearly see a gapless, linearly dispersing,
non-chiral charge-carrying edge mode throughout the FM phase. One can
interpret this as the
helical mode  of a strongly interacting
Luttinger liquid at the edge, an interpretation we will explore in
detail in future work. This mode shows up in the spectral densities of
both the $\rho\rho$ and the $S_zS_z$ correlators.  When one goes
through the transition into the CAF phase, the $\rho\rho$ correlator
should become gapped. We do see the gapped nature deep in the CAF
phase, but close to the transition, the finite size of our system
leads to some ``contamination'' from the gapless Goldstone mode of the
CAF bulk.

Many open questions remain. While the HF and TDHF approximations are
adequate far from the transition, we expect interaction corrections
beyond TDHF to play a role close to the transition. The bulk
transition is in the same universality class as the
Bose-Hubbard\cite{Bose-Hubbard} superfluid-insulator transition away
from the tip of the Mott lobes\cite{Ma-Halperin-Lee}. It has dynamical
critical exponent $z=2$, and at $T=0$ the interactions will be
marginally (but dangerously) irrelevant. Even more important is the
effect of these critical fluctuations on the charge-carrying modes at
the edge, and thus on the transport. Last, but not least, we have
assumed the system to be clean. Disorder could have a profound and
nonperturbative effect\cite{Bose-Hubbard} on the region near the phase
transition of the clean system. We hope to address these and other
questions in the near future.

Useful discussions with A. Young, P. Jarillo-Herrero, R. Shankar, and E. Berg
are gratefully acknowledged. The
authors thank the Aspen Center for Physics (NSF Grant No. 1066293) for
its hospitality, and for support by the Simons Foundation (ES). This
work was supported by the US-Israel Binational Science Foundation
(BSF) grant 2012120 (ES, GM, HAF), the Israel Science Foundation (ISF)
grant 231/14 (ES),  and NSF-DMR 1306897 (GM),
and by NSF-DMR 1506460.

\end{document}